\documentclass[10pt, journal]{IEEEtran}

\usepackage[ruled,linesnumbered,vlined]{algorithm2e}

\usepackage{amsmath,amsfonts}
\usepackage{array}
\usepackage{textcomp}
\usepackage{stfloats}
\usepackage{url}
\usepackage{verbatim}
\usepackage{graphicx}
\usepackage{cite}
\usepackage{listings}
\usepackage{xcolor}
\usepackage{tcolorbox}

\newcommand{\hll}{\colorbox[rgb]{0.631,0.780,0.55}}

\usepackage{subfigure}
\usepackage{amsmath}
\usepackage{amsthm}
\usepackage{graphicx} 
\usepackage{caption}
\usepackage{graphicx}
\usepackage{multicol} 
\usepackage{graphicx} 
\usepackage{amssymb}
\usepackage{float} 
\usepackage{wrapfig} 
\usepackage{booktabs}
\usepackage{lipsum} 
\usepackage{indentfirst}
\usepackage{bm}
\usepackage{color}
\usepackage{amsmath}
\usepackage{enumerate}
\usepackage{extarrows}
\usepackage{url}

\usepackage{graphicx}
\usepackage{epstopdf}
\usepackage{multirow}
\usepackage{tablefootnote}
\usepackage[flushleft]{threeparttable}
\usepackage{array}
\usepackage{amsmath}
\usepackage{enumitem}
\usepackage{enumerate}
\usepackage{tcolorbox}

\usepackage{makecell}
\usepackage{bbding}

\usepackage[colorlinks, linkcolor=blue, anchorcolor=blue, citecolor=blue]{hyperref}

\begin{document}

\definecolor{codegreen}{rgb}{0,0.6,0}
\definecolor{codegray}{rgb}{0.5,0.5,0.5}
\definecolor{codepurple}{rgb}{0.58,0,0.82}
\definecolor{backcolour}{rgb}{0.95,0.95,0.92}

\lstdefinestyle{mystyle}{
commentstyle= \color{red!50!green!50!blue!50},  
keywordstyle= \color{blue!70},  
numberstyle=\tiny\color{codegray},  
stringstyle=\color{codepurple},
basicstyle=\ttfamily\footnotesize,
breakatwhitespace=false,
breaklines=true,  
captionpos=b,
keepspaces=true,
numbers=left,  
numbersep=5pt,
showspaces=false,
showstringspaces=false,  
showtabs=false,
tabsize=2,
frame=single  
}

\title{OpenSN: An Open Source Library for Emulating LEO Satellite Networks}

\author{Wenhao~Lu,
Zhiyuan~Wang,
Hefan~Zhang,
Shan~Zhang,
and Hongbin~Luo
\IEEEcompsocitemizethanks{
\IEEEcompsocthanksitem Part of the results have been presented in APNet 2024 \cite{apnet2024}.
\IEEEcompsocthanksitem This work was supported by National Natural Science Foundation of China under Grants U24B20128, 62225201, 62202021, 62422201, 62271019, and in part by the National Key R\&D Program of China under Grant 2022YFB4501000, and in part by the Grant ANT2024003.
\textit{Zhiyuan Wang is the corresponding author}.
\IEEEcompsocthanksitem Wenhao Lu is with the School of Computer Science and Engineering, Beihang University, Beijing 100191, China. (Email: ZB2423318@buaa.edu.cn).
\IEEEcompsocthanksitem Zhiyuan Wang is with the School of Cyber Science and Technology, Beihang University, Beijing 100191, China, the School of Computer Science and Engineering, Beihang University, Beijing 100191, China, Zhongguancun Laboratory, Beijing 100191, China, and the State Key Laboratory of Virtual Reality Technology and Systems. (Email: zhiyuanwang@buaa.edu.cn).
\IEEEcompsocthanksitem Hefan Zhang is with School of Computer Science and Engineering, Beihang University, Beijing 100191, China. (Email: zhanghefan@buaa.edu.cn).
\IEEEcompsocthanksitem Shan Zhang and Hongbin Luo are with the School of Cyber Science and Technology, Beihang University, Beijing 100191, China, the School of Computer Science and Engineering, Beihang University, Beijing 100191, China, Zhongguancun Laboratory, Beijing 100191, China, and the State Key Laboratory of Software Development Environment. (Email: \{zhangshan18,luohb\}@buaa.edu.cn).
}	
}

\IEEEtitleabstractindextext{
\begin{abstract}
Low-earth-orbit (LEO) satellite constellations (e.g., Starlink) are becoming a necessary component of future Internet.
There have been increasing studies on LEO satellite networking.
It is a crucial problem how to evaluate these studies in a systematic and reproducible manner.
In this paper, we present OpenSN, i.e., an open source library for emulating large-scale satellite network (SN).
Different from Mininet-based SN emulators (e.g., LeoEM), OpenSN adopts container-based virtualization, thus allows for running distributed routing software on each node, and can achieve horizontal scalability via flexible multi-machine extension.
Compared to other container-based SN emulators (e.g., StarryNet), OpenSN streamlines the interaction with Docker command line interface and significantly reduces unnecessary operations of creating virtual links.
These modifications improve emulation efficiency and vertical scalability on a single machine.
Furthermore, OpenSN separates user-defined configuration from container network management via a Key-Value Database that records the necessary information for SN emulation.
Such a separation architecture enhances the function extensibility.
To sum up, OpenSN exhibits advantages in efficiency, scalability, and extensibility, thus is a valuable open source library that empowers research on LEO satellite networking.
Experiment results show that OpenSN constructs mega-constellations 5X-10X faster than StarryNet, and updates link state 2X-4X faster than LeoEM.
We also verify the scalability of OpenSN by successfully emulating the five-shell Starlink constellation with a total of 4408 satellites.
\end{abstract}

}

\maketitle

\IEEEdisplaynontitleabstractindextext
\IEEEpeerreviewmaketitle

\section{Introduction}
\subsection{Background \& Motivation}
In recent years, we have witnessed the rapid development of low-earth-orbit (LEO) constellations.
Different from the geostationary satellites orbiting at the altitude of 35768 km, the orbital altitude of LEO satellites is around 300-2000 km.
The lower orbital altitude leads to a smaller latency for the ground-satellite links (GSLs), but also reduces the coverage area of a single satellite.
Consequently, it is indispensable to deploy hundreds or thousands of LEO satellites to achieve global coverage.
Many commercial enterprises have launched their mega-constellation plans.
For example, Starlink Shell-I consists of a total of 1584 satellites on 72 orbital planes.
At the early stages, LEO satellites are only used as repeaters between two ground stations (GSes), which is also known as the bent-pipe architecture.
Recently, many LEO constellations have been equipped with (or are deploying) inter-satellite links (ISLs).
For example, Iridium Next has been using ISLs in L-band~\cite{IridiumISL}.
Starlink launched satellites equipped with laser-based ISLs way back in 2021~\cite{StarlinkISL}.
The adoption of ISLs creates the satellite network (SN) in space.

Although SN provides opportunities for extending Internet services to where the terrestrial networks cannot reach, it also leads to new challenges in satellite networking.
Due to the mobile nature and the time-varying communication links, the topology characteristics of LEO satellite networks are quite different from those of the terrestrial Internet.
There have been increasing studies on LEO satellite networking, focusing on the low-latency advantages (e.g.,~\cite{handley2018delay,bhattacherjee2018gearing,handley2019using,giuliari2020internet}), the constellation topology (e.g.,~\cite{bhattacherjee2019network,zhang2022enabling,li2021internet}), the shortest-path problem (e.g., \cite{chen2024shortest}), the intra-domain routing (e.g., host-centric~\cite{pan2019opspf,shan2023routing,taleb2008explicit}, content-centric~\cite{liang2021ndn,yan2024load,xia2021adapting}, and link-identified~\cite{zhang2024link,yan2024logic,zhang2023link}), the inter-domain routing (e.g., BGP-S~\cite{ekici2001network}, PCR~\cite{zeng2024adaptive}, DB-R~\cite{huang2024route}), and the security authentication (e.g., \cite{wang2024enabling}).

Despite the wide investigation, how to evaluate these new architectures, protocols, and algorithms in a systematic and reproducible manner has been an open problem.
Different from terrestrial Internet, it is costly to carry out experiments in a real-world SN.
Even for those commercial giants like SpaceX and Amazon, it also takes a few years to deploy their LEO constellations.
Therefore, most existing studies on SN still rely on network simulation or network emulation to evaluate the performance of their proposals.
There are significant differences between emulating LEO satellite networks and terrestrial networks. 
First of all, LEO constellations often consist of hundreds or thousands of satellites.
It is necessary to construct a large-scale topology for simulation/emulation.
Second, unlike terrestrial networks with fixed nodes, LEO satellites are in constant motion, leading to dynamic changes in visibility, connectivity, and propagation delays. 
This also increases the overhead of simulation/emulation.
In general, the simulation/emulation platforms for SN could be classified into three categories:

\textbf{(1) Orbit Analysis:}
The initial step of simulating SN is to obtain the trajectory data of LEO satellites.
STK \cite{stk} is an orbit analysis tool, but it does not support network protocol simulation.
Hence STK is often used as the orbit analyzer in this research area.

\textbf{(2) Discrete-Event Simulation:}
The discrete-event simulators of SN construct the constellation topology based on the trajectory data obtained from STK, and then simulate the networking events at different levels.
Specifically, the flow-level SN simulators (e.g., StarPerf \cite{lai2020starperf}) cannot capture the fine-grained packet forwarding behaviors, thus are not able to simulate networking protocols.
Packet-level SN simulators (e.g., \cite{kassing2020exploring,yan2023comparative,cheng2020comprehensive}) usually rely on a third-party discrete-event simulation platform (e.g., ns3~\cite{ns3} and OMNeT++~\cite{omnet}).
Typical networking protocols could be simulated based on packet-level SN simulators.

\textbf{(3) Virtual-Network Emulation:}
The virtual-network emulators of SN run real protocol stack in Linux kernel and support unmodified applications on host OS.
Existing virtual-network SN emulators either rely on Mininet \cite{mininet} or Docker container \cite{docker}.
For example, LeoEM \cite{cao2023satcp} is developed upon Mininet.
StarryNet \cite{lai2023starrynet} and the emulator in \cite{pan2022docker} adopt the container-based virtualization.

Among these solutions, virtual-network emulation provides the most realistic and meaningful results.
However, existing virtual-network SN emulators have some shortcomings.
For example, existing virtual-network SN emulators are not efficient enough to emulate frequent state changes (e.g., ISL failure/recovery and GSL handover), which slows down the experimental progress and prolongs the research period.
Moreover, existing virtual-network SN emulators often have scalability issues, making it difficult to emulate large-scale LEO constellations (e.g., Starlink).
Furthermore, most SN emulators are not fully open source, thus cannot provide a systematic evaluation platform that contributes to reproducibility.
In this paper, we would like to present OpenSN, i.e., an open source library for emulating large-scale satellite networks.
OpenSN is a virtual-network SN emulator that exhibits advantages in emulation efficiency, system stability, and function extensibility.
We believe that OpenSN could empower the research on LEO satellite networking in the future.
OpenSN has been available at \textsf{https://opensn-library.github.io}.

\subsection{Main Results and Key Contributions}

This paper presents OpenSN, i.e., an open source library for emulating LEO satellite networks.
Specifically, OpenSN is a virtual-network emulator, that runs real kernel and real applications.
Compared to other SN emulators (e.g., LeoEM \cite{cao2023satcp} and
StarryNet \cite{lai2023starrynet}), OpenSN achieves better emulation efficiency, system scalability, and function extensibility.
Next we introduce the key contributions and leave the detailed comparison (with previous works) in Table~\ref{Table: literature_2}.

\begin{itemize}
\item \textbf{Container-based Emulation:}
OpenSN adopts container-based virtualization for emulating SN with two considerations.
First, the container-based emulation achieves good horizontal scalability via flexible multi-machine extension.
Second, the container-based emulation allows for running distributed software (e.g., routing software and observation applications) in large-scale constellations without manual configuration.
The two aspects above enable OpenSN to outperform LeoEM (adopting lightweight virtualization via Mininet) in terms of horizontal scalability and function extensibility, respectively.

\item \textbf{Separation Architecture:}
OpenSN mainly consists of User-Defined Configurator, Container Network Manager, and Key-Value Database.
The Key-Value Database records the necessary information for SN emulation, and acts as a data forwarder between User-Defined Configurator and Container Network Manager.
Compared to other container-based SN emulators (e.g., StarryNet), such a separation architecture enhances the function extensibility of OpenSN.
For example, OpenSN allows users to flexibly change the emulation parameters and customize the emulation rules (e.g., GSL handover policy and ISL failure model) without complicated configuration.

\item \textbf{Efficiency Improvement:}
By investigating and comparing existing container-based SN emulators, we find that the emulation efficiency of the container network is mainly limited by the interaction with the Docker command line interface (CLI).
To this end, OpenSN streamlines the interaction process with Docker.
First, OpenSN's Container Runtime Manager replaces the CLI interaction with the official SDK, and also adopts a distributed architecture to aFchieve better parallelism.
Second, OpenSN's Virtual Link Manager skips Docker Network Manager, and directly controls Linux virtual devices, which can significantly reduce the overhead of unnecessary operations.
This way, OpenSN (installed distributed routing software) is still capable of achieving almost the same emulation efficiency as the specialized network emulator Mininet (without routing software).


\end{itemize}

The remainder of this paper is organized as follows.
We review related works on SN simulation/emulation in Section~\ref{Section: Literture Review}.
We present the OpenSN architecture in Section~\ref{Section: Framework}.
We introduce the usage of OpenSN in Section~\ref{Section: Usage}.
We evaluate the performance of OpenSN in Section~\ref{Section: Evaluation}.
Finally, we conclude this paper in Section~\ref{Section: Conclusion}.

\begin{table*}[t]
\renewcommand{\arraystretch}{1.4}
\centering	
\caption{Network simulators/emulators} 
\label{Table: literature_all}
{\small
\begin{tabular}{|c|c|c|c|c|c|c|}
\hline
\textbf{Category} & \textbf{Tools} & \textbf{Granularity} & \textbf{Technology} & \textbf{Networking} & \textbf{Open Source} \\ 
\hline\hline
\multirow{4}{*}{\makecell[c]{Discrete-\\event\\simulator}}& StarPerf~\cite{lai2020starperf} & Flow-level  & STK~\cite{stk} \& Python \& Matlab & Route calculation & $\checkmark$ \\
\cline{2-6}
\cline{2-2}\cline{4-6}
& Hypatia~\cite{kassing2020exploring}	& \multirow{3}{*}{Packet-level} & ns3 \cite{ns3} & \multirow{3}{*}{\makecell[c]{Routing protocol\\simulation}} & $\checkmark$ \\
\cline{2-2}\cline{4-4}\cline{6-6}
& Yan \textit{et al.} in \cite{yan2023comparative} & & OMNeT++~\cite{omnet} &  & $\times$ \\
\cline{2-2}\cline{4-4}\cline{6-6}
& Cheng \textit{et al.} in \cite{cheng2020comprehensive} & & ns3~\cite{ns3} \& Matlab & & $\times$\\
\hline
\multirow{9}{*}{\makecell[c]{Virtual-\\network\\emulator}} & Mininet~\cite{mininet}	& \multirow{9}{*}{Real stacks} &  Virtualization (Namespace) & \multirow{2}{*}{NA}  & \multirow{2}{*}{Not for SN}\\
\cline{2-2}\cline{4-4}
& Etalon~\cite{mukerjee2020adapting}	&  &  Container (Docker) &  &  \\
\cline{2-2}\cline{4-6}
& LeoEM~\cite{cao2023satcp} &  & Mininet~\cite{mininet} & Route calculation & $\checkmark$ \\
\cline{2-2}\cline{4-6}
& StarryNet~\cite{lai2023starrynet} &  & Container (Docker) & \multirow{4}{*}{\makecell[c]{Routing software\\in containers/VMs}} & Partially \\
\cline{2-2}\cline{4-4}\cline{6-6}
& Pan \textit{et al.} in~\cite{pan2022docker} &  & Container (Docker) &  & $\times$ \\
\cline{2-2}\cline{4-4}\cline{6-6}
& Celestial~\cite{pfandzelter2022celestial} &  & MicroVM (firecracker~\cite{agache2020firecracker}) &  & $\checkmark$ \\
\cline{2-2}\cline{4-4}\cline{6-6}
& NEaaS~\cite{lai2021network} &  & Virtual Machine (KVM~\cite{kivity2007kvm}) &  & $\times$ \\
\cline{2-2}\cline{4-4}\cline{4-6}\cline{6-6}
& LORSAT~\cite{afhamisis2022testbed}  &  & Virtual Machine (VMWare) & NA & $\times$ \\
\cline{2-2}\cline{4-4}\cline{4-6}\cline{6-6}
& \textbf{OpenSN} &  & Container (\textit{Better Manager}) & Routing software in containers  & $\checkmark$ \\
\hline
\end{tabular}
}
\end{table*}

\begin{table*}[t]
\renewcommand{\arraystretch}{1.6}
\centering
\caption{Performance comparison of SN emulators} 
\label{Table: literature_2}
{\small
\begin{tabular}{|c|c|c|c|c|c|}
\hline
\textbf{SN Emulator} & \textbf{Emulation Efficiency} & \textbf{System Scalability} & \textbf{Function Extensibility}\\ 
\hline\hline
LeoEM& \makecell[c]{Efficient for SN emulation\\thanks to the lightweight\\virtualization of Mininet} & \makecell[c]{Good vertical scalability on\\single machine, not horizontally\\ scalable to more machines} & \makecell[c]{Not able to emulate SN running\\distributed software (e.g., routing)\\due to the lightweight virtualization}\\
\hline
StarryNet & \makecell[c]{Inefficient due to direct\\interaction with Docker\\CLI in SN emulation} & \makecell[c]{Good horizontal scalability to\\multiple machines, but is not\\vertically scalable due to direct\\interaction with Docker CLI} & \makecell[c]{Allow for distributed software\\(e.g, routing), but not flexible\\ to modify images, topology, etc} \\
\hline
OpenSN & \makecell[c]{Efficient due to our\\improvement on the\\process of using Docker\\CLI in our managers}  & \makecell[c]{Good vertical and horizontal\\scalability due to improvement\\on the process of using Docker\\CLI in SN emulation} & \makecell[c]{Allow for distributed software\\(e.g, routing), and flexible to extend\\(due to separation of user configuration\\from container network management)} \\
\hline
\end{tabular}
}
\end{table*}

\section{Literature Review}
\label{Section: Literture Review}

The existing platforms for evaluating satellite networking protocols could be classified into two categories.
We will compare their advantages and disadvantages in the following.
Table~\ref{Table: literature_all} provides a guideline.

\subsection{Discrete-Event Simulation for SN}
The discrete-event simulation for SN captures the networking events at different levels:

\textit{Flow-level:}
The flow-level SN simulators could calculate some performance metrics (e.g., end-to-end propagation delay) given the constellation topology.
However, they are not able to capture the fine-grained behaviors of packets under different networking protocols.
Lai \textit{et al.} in~\cite{lai2020starperf} present StarPerf, i.e., a flow-level SN simulator based on Python.
It allows for trajectory simulation, topology generation, and shortest-path calculation.
They leverage StarPerf to gain insights
on optimizing the constellation architecture.

\textit{Packet-Level:}
Existing packet-level SN simulators rely on the third-party discrete-event simulation platforms, e.g., ns3 \cite{ns3} and OMNeT++ \cite{omnet}.
Kassing \textit{et al.} in \cite{kassing2020exploring} propose Hypatia, i.e., a packet-level SN simulator based on ns3.
The authors use Hypatia to investigate the impact of latency and link utilization on congestion control and routing.
Cheng \textit{et al.} in \cite{cheng2020comprehensive} present a packet-level simulator for space-air-ground integrated network based on ns3, which supports various mobility traces of space and aerial networks.
Both the two SN simulators above focus on the classic TCP/IP architecture.
Yan \textit{et al.} in \cite{yan2023comparative} build a packet-level SN simulator on OMNET++, and compare the performance between IP architecture and Named Date Network(NDN) architecture in the LEO satellite network scenario.
The authors further conduct a comparative study of host-centric routing and information-centric routing in large-scale LEO satellite constellations.

The aforementioned packet-level SN simulators have two drawbacks.
First, it usually takes a long time to complete a simulation run due to the complexity of scheduling discrete events.
For example, when the LEO constellation is large, OSPF protocol generates a great number of message exchanges and link-state announcements, as well as the routing table calculation events.
These discrete events are not able to be scheduled in a parallel manner.
Second, they are not able to run real protocol stacks of OS.
This means that the packet-level simulation results still exhibit differences compared to the real-world system.
Therefore, some studies have been exploring the virtual-network emulation for SN.

\subsection{Virtual-Network Emulation for SN}
The virtual-network emulators for SN mainly depend on three types of virtualization technologies:
(i) Mininet (trimmed process virtualization); 
(ii) Container technology; 
(iii) Virtual machine technology.
One could run real applications over SN with real-time response.

\textbf{Mininet-based:}
Mininet~\cite{mininet} enables users to create a realistic virtual network, running a real kernel stack.
However, Mininet does not provide specific support for SN emulation.
Moreover, Mininet adopts lightweight virtualization (i.e., process-based) \cite{mininet}, thus is not suitable to emulate large-scale SN running distributed software (e.g., routing software).
Cao \textit{et al.} in~\cite{cao2023satcp} develop LeoEM (available at \cite{LeoEMURL}), which emulates LEO constellations via Mininet.
Moreover, LeoEM adopts StarPerf's solution for constellation construction and the shortest-path calculation.
The authors in~\cite{cao2023satcp} leverage LeoEM to evaluate their proposed congestion control mechanisms (i.e., SatTCP), aiming to address the challenge caused by frequent GSL handovers.

\textbf{Container-based:}
The container-based network emulation has also been adopted in different scenarios other than SN.
For example, Mukerjee \textit{et al.} in \cite{mukerjee2020adapting} present the design of Etalon (available at \cite{Etalon}), i.e., an open source reconfigurable datacenter network emulator developed upon Docker container.
The authors establish a testbed on four servers to emulate a datacenter network.
However, Etalon does not support for SN emulation.
Lai \textit{et al.} in~\cite{lai2023starrynet} develop StarryNet, i.e., a container-based emulator for SN.
It is also implemented upon Docker, and emulates the distributed routing process by loading BIRD \cite{bird} on each container node.
Particularly, StarryNet also incorporates some real orbital data (which is not the scope of our study in this paper).
StarryNet is now partially available at \cite{StarryNetURL}.
Furthermore, Pan \textit{et al.} in~\cite{pan2022docker} introduce a similar container-based SN emulator, using another routing software Quagga \cite{quagga}.
However, they do not open source this work.
Hence we are not able to provide performance comparison with it.

\textbf{VM-based:}
Virtual machine (VM) technology enables users to create and connect multiple isolated virtual network nodes on a single physical machine, forming a virtual network. 
There are also several VM-based SN network emulators.  
For example, LORSAT~\cite{afhamisis2022testbed} creates an emulation-based testbed integrating emulated satellite components with real LoRaWAN devices but only supports the bent-pipe architecture without routing calculation.
NEaaS~\cite{lai2021network} is a cloud-based network emulation platform offering Network Emulation as a Service. 
It uses VM-Container hybrid architecture to build the virtual network. 
However, virtual machine nodes still hold a dominant position in their virtual networks. 
The high overhead of virtual machines prevents large-scale (thousands of nodes) network emulation. 
Celestial~\cite{pfandzelter2022celestial} is a micro-VM-based \cite{agache2020firecracker} virtual testbed for the LEO edge, primarily designed for edge computing scenarios.
It create VMs only for involved satellites and lacks support for distributed routing emulation.
Although VMs offer better isolation and the capacity to customize the kernel for each network node, the excessive overhead makes these emulation tools unable to meet the emulation requirements of large-scale LEO satellite networks.

OpenSN is a container-based emulator for SN.
Compared to LeoEM and StarryNet, OpenSN exhibits advantages in emulation efficiency, system scalability, and function extensibility.
We briefly summarize the comparison in Table~\ref{Table: literature_2}, and will introduce the detailed design in Section~\ref{Section: Framework}.

\begin{figure*}
\centering
\includegraphics[width=0.9\linewidth]{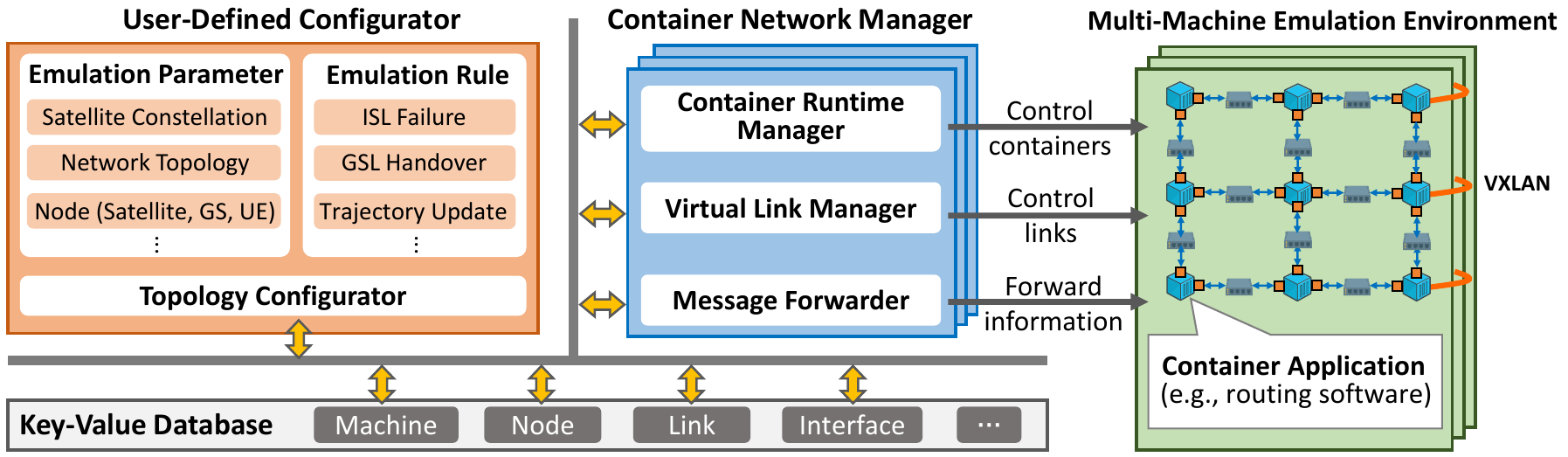}
\caption{System architecture of OpenSN.}
\label{fig: Architecture}
\end{figure*}

\section{OpenSN Framework}
\label{Section: Framework}

This section presents the design of OpenSN.
We first provide an overview in Section~\ref{Subsection: Overview}.
We then introduce the improvements on emulation efficiency and system scalability in Section~\ref{Subsection: Network Emulation} and Section~\ref{Subsection: Multi-Machine Extension}, respectively.
We introduce the enhancement via eBPF links in Section~\ref{Subsection: eBPF Link}.
Finally, we present the function extensibility of OpenSN in Section~\ref{Subsection: Extensibility}.

\subsection{Overview and Architecture}
\label{Subsection: Overview}

OpenSN is a virtual-network emulator for LEO satellite networks, which leverages containers and virtual links to emulate SN.
Fig.~\ref{fig: Architecture} shows the architecture of OpenSN, including three key components: User-Defined Configurator, Container Network Manager, and Key-Value Database.

\textbf{User-Defined Configurator} allows users to specify the emulation parameters (e.g., constellation, connectivity, nodes, etc) and the emulation rules (e.g., ISL failure model, GSL handover policy, trajectory udpate, etc) in OpenSN.
To enhance extensibility, the corresponding configuration data will not be directly passed to Container Network Manager, but will be recorded in Key-Value Database.

\textbf{Key-Value Database} records the necessary information of SN emulation.
Some are user-specified emulation parameters and rules from User-Defined Configurator.
Some are status data of SN emulation obtained from Container Network Manager in each machine.
Table~\ref{Table: KV Database} summarizes the information recorded in OpenSN's Key-Value Database, which includes the configuration for machines, nodes, links, and applications.
Specifically, each configuration may include required ones and user-defined ones.
\begin{itemize}
    \item {Required configurations} cover the necessary parameters or rules required by SN emulation. 
    For example, a machine's maximum number of containers specifies how many containers can run on each machine.
    Geometric parameters of satellites and ground stations (GS) are used to calculate the trajectories of satellites and control the ground-satellite link (GSL) handover and inter-satellite link (ISL) failure.
    
    \item User-defined configurations are the optional configurations that could be customized by OpenSN users. 
    These configurations will take effect in OpenSN User-Defined Configurator and container applications.
    In Table~\ref{Table: KV Database}, we list some examples used in Section~\ref{Section: Usage}.
    For example, we specify the \textit{timestamps for terminal user (TU) node to send requests to the content provider (CP)}.
    Furthermore, OpenSN leaves more possibilities to users to achieve a better extensibility.
\end{itemize}

\begin{table}
\renewcommand{\arraystretch}{1.1}	
\centering	
\caption{Information of Key-Value Database} 
\label{Table: KV Database}
\begin{tabular}{|c|l|c|c|c|c|c|}
\hline
\textbf{Type} & $\qquad\qquad\qquad\qquad$\textbf{Parameter}\\ 
\hline\hline
\multirow{3}{*}{\makecell[c]{Machine\\(\textit{Required})}} & Maximal number of containers\\
\cline{2-2}
& IP address of network interface card \\
\cline{2-2}
& MAC address of network interface card \\
\hline
\multirow{5}{*}{\makecell[c]{Node\\(\textit{Required})}} & 
Node type (i.e., satellite, GS, TU, router) \\
\cline{2-2}
& ID of links maintained by the node\\
\cline{2-2}
& Index of machine for this node \\
\cline{2-2}
& Geometric parameters of satellite and GS nodes \\
\cline{2-2}
& Orbit Index of satellite nodes \\
\hline
\multirow{3}{*}{\makecell[c]{Node\\(\textit{User-defined})}}& Gateway addresses of TU and CP nodes \\
\cline{2-2}
& File paths that should be exposed in the node \\
\cline{2-2}
& \textit{OpenSN leaves more possibilities to users} \\
\hline
\multirow{3}{*}{\makecell[c]{Link\\(\textit{Required})}} & Link address (e.g., IPv4/IPv6 address or others) \\
\cline{2-2}
& Link parameter (i.e., distance, delay, bandwidth) \\
\cline{2-2}
& Type, i.e., GSL and inter-orbit/inner-orbit ISL \\
\hline
\multirow{3}{*}{\makecell[c]{Link\\(\textit{User-defined})}}& 
Maximum transmission unit \\
\cline{2-2}
& Whether to be monitored \\
\cline{2-2}
& \textit{OpenSN leaves more possibilities to users} \\
\hline
\multirow{2}{*}{\makecell[c]{Application\\(\textit{Required})}}& Timestamp to be launched \\
\cline{2-2}
& Autonomous system index of router \\
\hline
\multirow{3}{*}{\makecell[c]{Application\\(\textit{User-defined})}}& Timestamps for TU node to send requests \\
\cline{2-2}
& Experimental results recorded metrics options \\
\cline{2-2}
& \textit{OpenSN leaves more possibilities to users} \\
\hline
\end{tabular}
\end{table}

\textbf{Container Network Manager} controls the major functionality of SN emulation, which is conducted on multiple machines connected by VXLAN.
According to the information from Key-Value Database, Container Runtime Manager is responsible for container creation/deconstruction and single-node resource management.
Virtual Link Manager constructs the virtual links for each ISL and GSL in SN.
Moreover, Message Forwarder is used to transfer necessary information (e.g., real-time topology, IP address of each interface, and user's command) between Key-Value Database and Multi-Machine Emulation Environment.

\subsection{Efficient Network Emulation}
\label{Subsection: Network Emulation}
\begin{figure}
\centering
\includegraphics[width=1\linewidth]{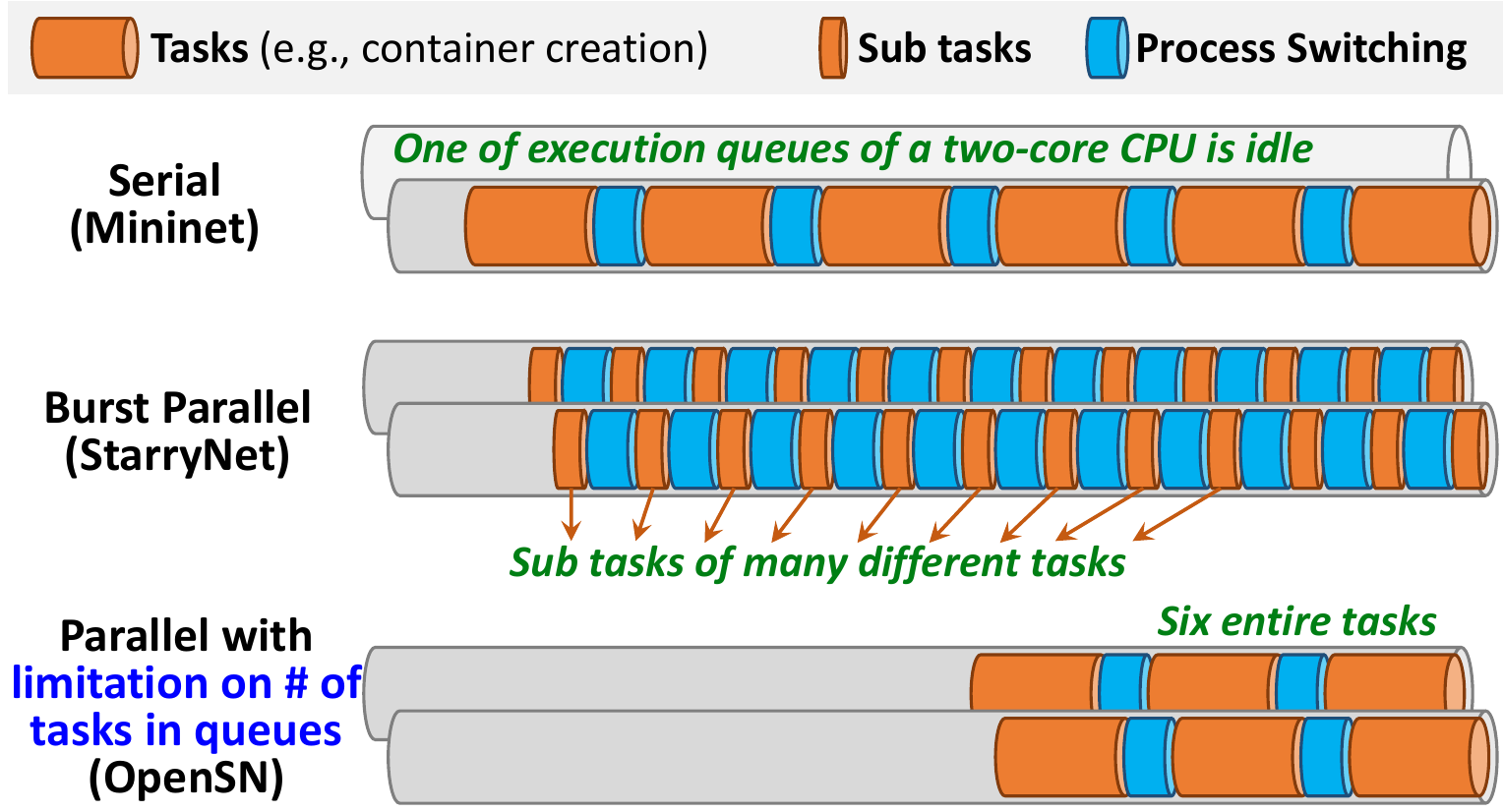}
\caption{Comparison of task scheduling}
\label{fig: task-sched-cmp}
\end{figure}

One of the major challenges and performance bottlenecks of emulating SN lies in the network emulation.
On the one hand, the number of involved nodes, ISLs and GSLs is huge for LEO mega-constellations.
On the other hand, the link state changes (e.g., ISL failure/recovery and GSL handover) are quite frequent due to the mobile nature of LEO satellites.
These two aspects above require an efficient network manager.
Existing container-based SN emulators are not efficient enough due to the following two drawbacks.
\begin{itemize}
\item The Docker network controller is not developed for emulating large-scale networks with frequent state changes.
Specifically, the Docker network manager exhibits unnecessary processes in container management and link management (e.g., repeatedly interacting with data stores).
Moreover, concurrent network state changes (e.g., creation/deconstruction of containers and links) may be blocked by the Docker Daemon in the same task queue, which reduces the response speed of network emulation.
This becomes an even more significant overhead when the constellation scales up.

\item The existing SN emulators do not optimize the scheduling of network construction tasks.
Roughly speaking, the creation of a container network includes two steps: node creation (heavy computation) and link creation (light computation).
As shown in Fig.~\ref{fig: task-sched-cmp}, serial execution (e.g., Mininet) will lead to insufficient utilization of CPU resources for light-computation tasks, thus eventually slowing down the network construction.
Parallel execution (e.g., StarryNet) will cause frequent process switching and blocking for computation-intensive tasks, thus eventually slowing down the network construction.
\end{itemize}
Based on the above observations, Container Network Manager tends to improve the emulation efficiency from the perspectives of nodes and links.
We first introduce how OpenSN streamlines the procedure of node management and link management via Container Runtime Manager and Virtual Link Manager, respectively.
We then introduce their interactions.

\textbf{Container Runtime Manager} of OpenSN is implemented in a distributed manner and should be deployed on each machine.
To improve the efficiency, we make the following improvements.
First, Container Runtime Managers of OpenSN in each machine rely on a coroutine pool to schedule the node creation/deconstruction tasks in a parallel manner. 
The capacity of such a coroutine pool is determined by the CPU specification of the machine, which can limit the number of simultaneous tasks and avoid excessive process switching.
Second, Container Runtime Manager of OpenSN leverages the official Docker SDK to connect with Docker daemon, instead of relying on Docker Command Line Interface (CLI) like StarryNet.
This enables OpenSN to skip many unnecessary processes (e.g., launch shell and Docker-client) when creating containers or executing commands inside containers, thus significantly reducing the operation overhead.

\begin{figure}
\centering
\includegraphics[width=1\linewidth]{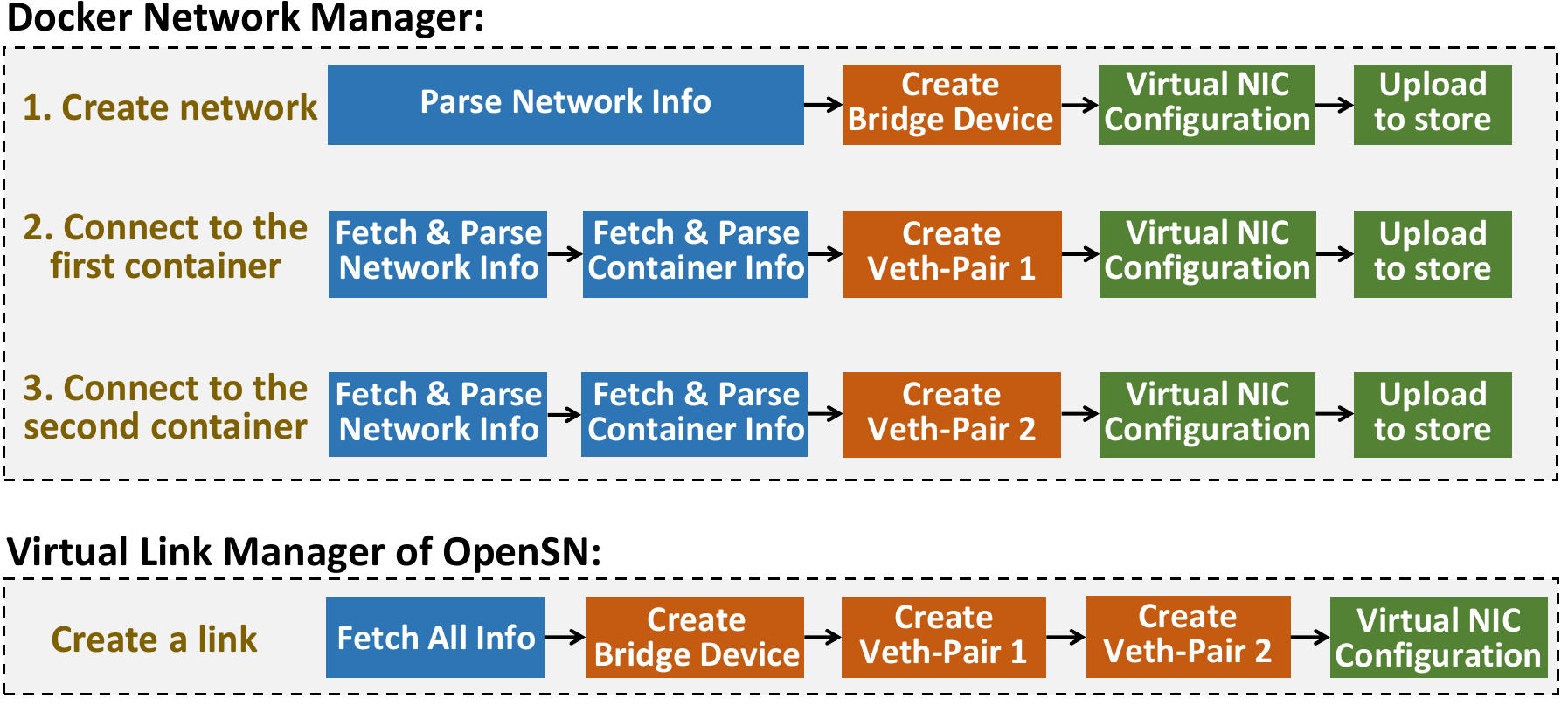}
\caption{Process of creating a link under Docker Network Manager and OpenSN's Virtual Link Manager}
\label{fig: vlink-create-cmp}
\end{figure}

\textbf{Virtual Link Manager} of OpenSN is developed to directly manage the Linux virtual network devices and the network namespace. 
This manager has the highest priority.
It will immediately create a coroutine to implement rapid response once the command of link-state change is detected. 
Furthermore, OpenSN's Virtual Link Manager streamlines the process of creating a virtual link.
Fig.~\ref{fig: vlink-create-cmp} shows the event sequence of creating a virtual link under Docker Network Manager and OpenSN's Virtual Link Manager.
Specifically, Docker Network Manager creates a link in three major actions:
\begin{itemize}
\item Create network,
\item Connect to the first container,
\item Connect to the second container.
\end{itemize}
Moreover, each action actually consists of several steps, e.g., parsing/fetching information and other necessary configurations.
Such a process will reduce efficiency of link operations.
For example, the interaction with a data store will trigger concurrent synchronization throttling, reducing the parallelism of task execution.
In contrast, Virtual Link Manager of OpenSN could establish a virtual link within a single call, which improves the parallelization efficiency of link operations.

\textbf{Interaction between Node \& Link Creation:}
The interaction between node creation and link creation also affects the emulation efficiency.
Roughly speaking, link creation tasks cannot be executed until the corresponding two containers have been created.
StarryNet and Mininet start to execute link creation tasks after all the containers in the emulation environment have been created. 
Obviously, such a scheduling pattern is not efficient in terms of time.
OpenSN tends to optimize the interaction between node creation and link creation, aiming to speed up the network construction progress.
To this end, OpenSN implements a fine-grained orchestrator for dependent tasks, and aims to create correlated containers and links together.
Specifically, Virtual Link Manager of OpenSN has a waiting pool, which records all the received link creation tasks.
Upon a container being created, Container Runtime Manager of OpenSN would signal the waiting pool.
Virtual Link Manager then starts to create links of this container.


Although OpenSN has improved the efficiency of link creation, it is still a significant overhead to configure a great number of link handovers at one time.\footnote{Here link handover means that a container disconnects from container 1 and immediately connects to container 2.}
However, GSL handover is a crucial aspect in SN emulation.
To address this issue, OpenSN tends to improve the efficiency of emulating GSL handover via eBPF.
In the following, we first present multi-machine extension in Section~\ref{Subsection: Multi-Machine Extension}.
We then introduce how to improve the handover efficiency for intra-machine links and inter-machine links in Section~\ref{Subsection: eBPF Link}.

\subsection{Multi-Machine Extension}
\label{Subsection: Multi-Machine Extension}
The virtual-network emulation for LEO mega-constellation requires a lot of computing resources.
Although OpenSN has made significant improvements in emulation efficiency in Section~\ref{Subsection: Network Emulation}, the pattern of single-machine deployment still cannot afford to emulate thousands of satellites running resource-hungry applications.
Hence OpenSN provides the capability of multi-machine extension.
Next we introduce how the control plane and data plane work.

\subsubsection{Control Plane}
\begin{figure}
\centering
\includegraphics[width=0.99\linewidth]{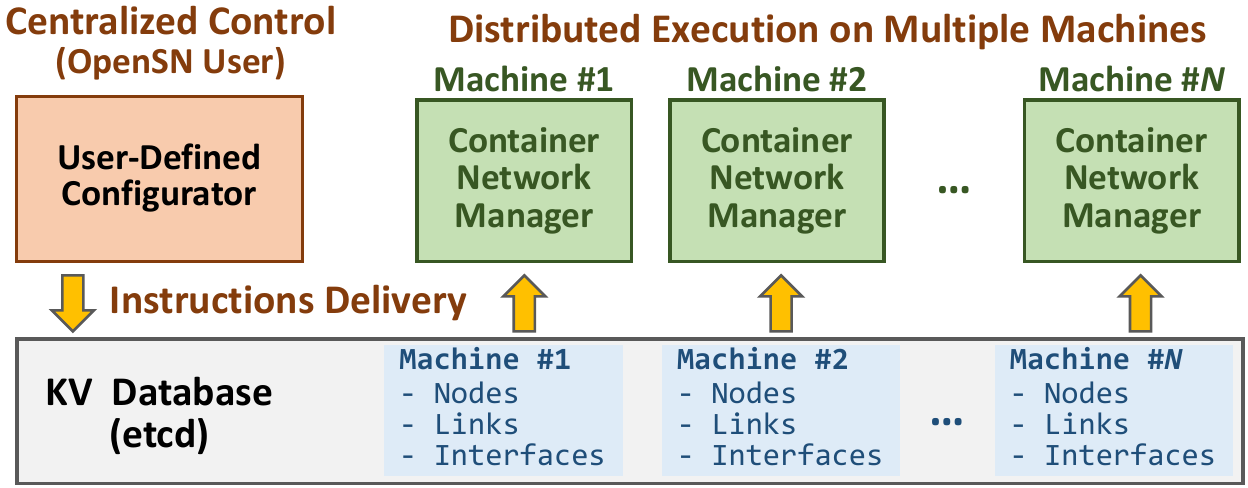}
\caption{Control plane framework for multi-machine extension.}
\label{fig: multimachine-control}
\end{figure}

To achieve flexible multi-machine extension, OpenSN's control plane consists of centralized control and distributed execution (in multiple machines).
Fig.~\ref{fig: multimachine-control} plots the control plane framework for multi-machine extension.

\textbf{Centralized Control} is achieved by User-Defined Configurator shown in Fig.~\ref{fig: Architecture}.
The OpenSN SDK for Python for User-Defined Configurator generates the detailed information (i.e., machine index, nodes, links, interfaces, etc) for each machine.
This process is transparent to OpenSN users.
This enables OpenSN users to control the configuration on a single machine without considering the entire emulation environment.

\textbf{Distributed Execution} refers to container network creation and operation on different machines. 
Each machine is responsible for emulating part of the LEO satellites and the corresponding links.
Such a parallel execution paradigm helps reduce the runtime resource consumption on a single machine, thus improving the execution efficiency. 
For load-balancing, we use a weighted round-robin procedure to assign containers and links to the machines. 
Algorithm~\ref{alg: weighted round robin} summarizes the procedure.
Algorithm~\ref{alg: weighted round robin} mainly consists of two phases. 
\begin{itemize}
    \item Lines~\ref{Line: Init Machines' Weight List}-\ref{Line: Init Machines' Weight List End} initialize the $\textit{WeightList}$ and the $\textit{WeightLeftList}$ according to the weight of machines. 
    The $\textit{WeightLeftList}$ is used to assign instances to each machine, and the $\textit{WeightList}$ is used to refresh the $\textit{WeightLeftList}$ in each round.
    This phase also initializes the variable $i$, which is the index of the current machine that the scheduler assigns instances to.
    \item Lines~\ref{Line: Assign Instances and Links}-\ref{Line: Assign Instances and Links End} assign the instances and links to the machines. 
    When an instance and its links have been assigned to the $\textit{Set}_{\textit{machine}}[i]$, the $\textit{WeightLeftList}[i]$ decreases.
    When the $\textit{WeightLeftList}[i]$ reaches zero, the $\textit{WeightLeftList}[i]$ will be refreshed.
    This loop will continue until all instances have been assigned.
\end{itemize}
At the end, the instances and links will be divided into several subsets and assigned to each machine according to the resource capacity of each machine.

\begin{algorithm}[t]
    \caption{Weighted Round Robin Scheduler}
    \label{alg: weighted round robin}
    
    \KwIn{$\textit{Set}_{\textit{machine}}$, $\textit{Set}_{\textit{instance}}$, $\textit{Set}_{\textit{link}}$}
    \KwOut{$\textit{Map}_{\textit{[machine,link]}}$, $\textit{Map}_{\textit{[machine,instance]}}$}
    
    \tcp{\hll{Initialize weight list}}
    \label{Line: Init Machines' Weight List}
    
    $\textit{WeightList} \leftarrow []$\\
    
    $\textit{WeightLeftList} \leftarrow []$\\
    
    \For{$\textit{machine} \in \textit{Set}_{\textit{machine}}$}{
        $\textit{WeightList.append}(\textit{machine.weight})$\\
        $\textit{WeightLeftList.append}(\textit{machine.weight})$\\
    }
    
    $i \leftarrow 0$  \label{Line: Init Machines' Weight List End}\\
    
    \tcp{\hll{Assign instances \& links}}
    
    \For{$\textit{instance} \in \textit{Set}_{\textit{instance}}$}{
    \label{Line: Assign Instances and Links}
        $\textit{Map}_{\textit{instance}}$[$\textit{Set}_{\textit{machine}}$[$i$]].append($\textit{instance}$)\\
        
        \For{ $\textit{link} \in \textit{Set}_{\textit{link}} \cap \textit{instance}.\textit{links}$ }{
            $\textit{Map}_{\textit{link}}[\textit{Set}_{\textit{machine}}[i]].\textit{append}(\textit{link}$)
        }
        $\textit{WeightLeftList}[i] \leftarrow \textit{WeightLeftList}[i] -1$\\
        
        \If{$\textit{WeightLeftList}[i] \leq 0$}{
            
            $\textit{WeightLeftList}[i] \leftarrow \textit{WeightList}[i]$\\
            
            $i \leftarrow i + 1 $\\
            
        }
    }
    \label{Line: Assign Instances and Links End}
    \textbf{return} $\textit{Target}$, $\textit{Map}_{\textit{link}}$, $\textit{Map}_{\textit{instance}}$
\end{algorithm}

\textbf{Instruction Delivery} is achieved by a KV database based on Etcd \cite{etcd}.
Specifically, Etcd records the status of different entities (e.g., containers and links).
For example, the status of a link entity could be connected or disconnected.
The instructions generated based on user configuration will be dispatched to the corresponding machines.
With the server-pushing ability, Etcd is able to immediately notify the Container Network Manager of the corresponding machine (once the status change is detected).
Accordingly, Container Network Manager will update the emulation environment.
This enables Container Network Manager of each machine to immediately respond to the frequent status changes (e.g., link delay and link connection) at a low cost in LEO constellations.

\subsubsection{Data Plane}
To achieve multi-machine extension, the data plane of OpenSN should be able to emulate the ISLs and GSLs when the corresponding containers are deployed on different emulation machines.
OpenSN leverages VXLAN technology \cite{mahalingam2014virtual} to achieve inter-machine data transfer. 
This is because most links of SN are point-to-point, and VXLAN naturally supports such a network structure and can flexibly connect containers across machines.
As we will see in Section~\ref{Subsection: eBPF Link}, 
OpenSN can further improve the emulation efficiency of inter-machine links via eBPF technique \cite{vieira2020fast}.

\begin{figure}
\centering
\includegraphics[width=0.95\linewidth]{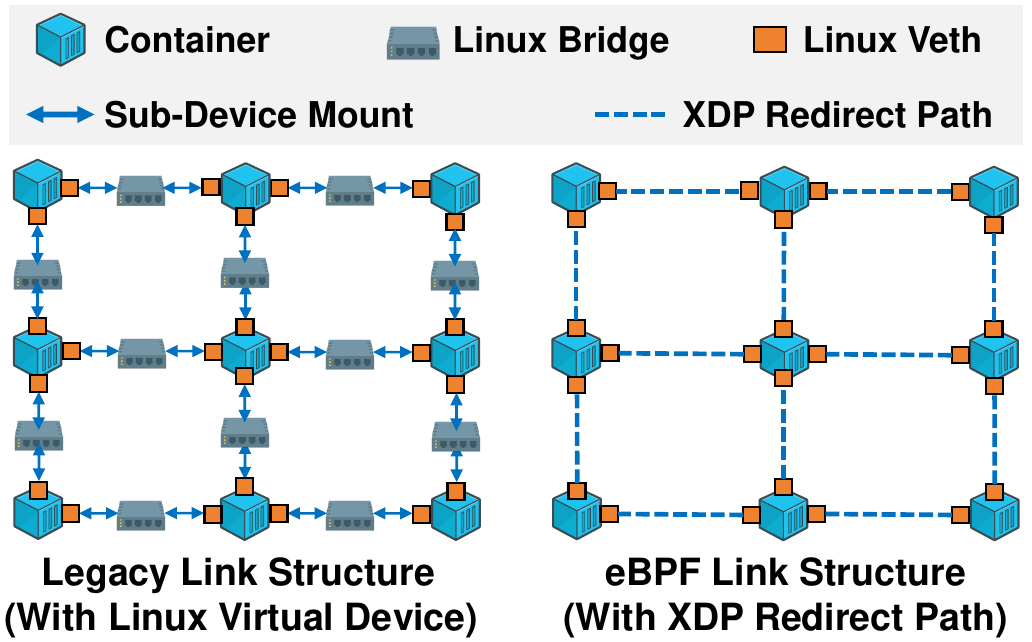}
\caption{Difference between legacy link and eBPF link}
\label{fig: diff-link}
\end{figure}

\subsection{Enhancement via eBPF Links}
\label{Subsection: eBPF Link}
We have introduced how OpenSN improves the emulation efficiency and scalability in Section~\ref{Subsection: Network Emulation} and Section~\ref{Subsection: Multi-Machine Extension}, respectively.
To further enhance the two aspects above, OpenSN incorporates a novel virtual link structure based on XDP (eXpress Data Path) \cite{vieira2020fast}.
This is especially beneficial to emulate frequent GSL handover and speed up the inter-machine data stream.
Next we first introduce eBPF links.
We then present intra-machine eBPF links and inter-machine eBPF links.

\subsubsection{eBPF Links}
The eBPF link is a virtual link structure, which uses eBPF program (attached to the XDP hook point of the virtual network device) to realize the function of redirecting frames.
Specifically, the eBPF program provides the ability to revise the kernel space data path.
With this capability, the eBPF link could outperform the traditional virtual links in the following two aspects.
For intra-machine links, eBPF links could replace all the Linux bridge devices with a redirect controller.
For inter-machine links, eBPF links could transmit the Ethernet frames by directly modifying destination MAC address instead of repacking the packets.
The two aspects above significantly reduce the number of Linux virtual devices, and streamline the operation of link handover. 

Next we introduce the detailed implementation of eBPF links for inner-machine and inter-machine cases, respectively.

\subsubsection{Inner-Machine eBPF Links}
We demonstrate the difference between eBPF link structure and legacy link structure based on Fig.~\ref{fig: diff-link}.
Specifically, the legacy link structure relies on Linux bridge devices.
In contrast, eBPF link structure uses the kernel-space eBPF program and eBPF map to implement frame redirection for virtual links.
Overall, eBPF link structure exhibits the following two-fold advantages:

\begin{figure}
\centering
\includegraphics[width=0.95\linewidth]{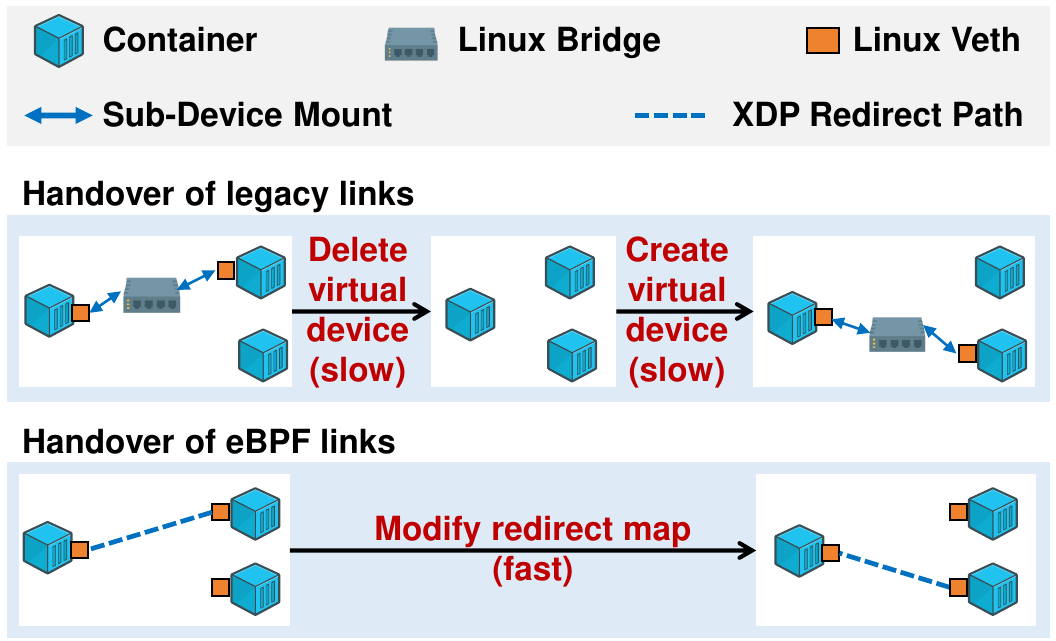}
\caption{Handover difference between legacy and eBPF links}
\label{fig: diff-switch}
\end{figure}

\textbf{Efficient Network Construction:}
As mentioned above, the process of creating an eBPF link skips the bridge device creation compared to the legacy link.
This reduces the burden of Virtual Link Manager, and facilitates efficient network construction and deconstruction.

\textbf{Better Runtime Efficiency:}
The eBPF technique also improves the runtime efficiency in terms of configuring link handover.
As shown in Fig.~\ref{fig: diff-switch}, the handover of legacy links involves deleting and recreating virtual devices (i.e., Linux bridges).
Such a process will inevitably change the interface inside the container, causing extra overhead.
Moreover, some switch functions (e.g., broadcast and port redirect map)  in Linux bridges are redundant for point-to-point links, which consume more computation resources than eBPF redirect program. 
In contrast, the handover of eBPF links can be achieved by directly modifying the redirect map, which is much faster than virtual device configuration.
Moreover, the container can keep the same interface after handover.
Therefore, eBPF technique enables OpenSN to efficiently change the connections of network nodes.

\subsubsection{Inter-Machine eBPF Links}
\begin{figure}
\centering
\includegraphics[width=0.95\linewidth]{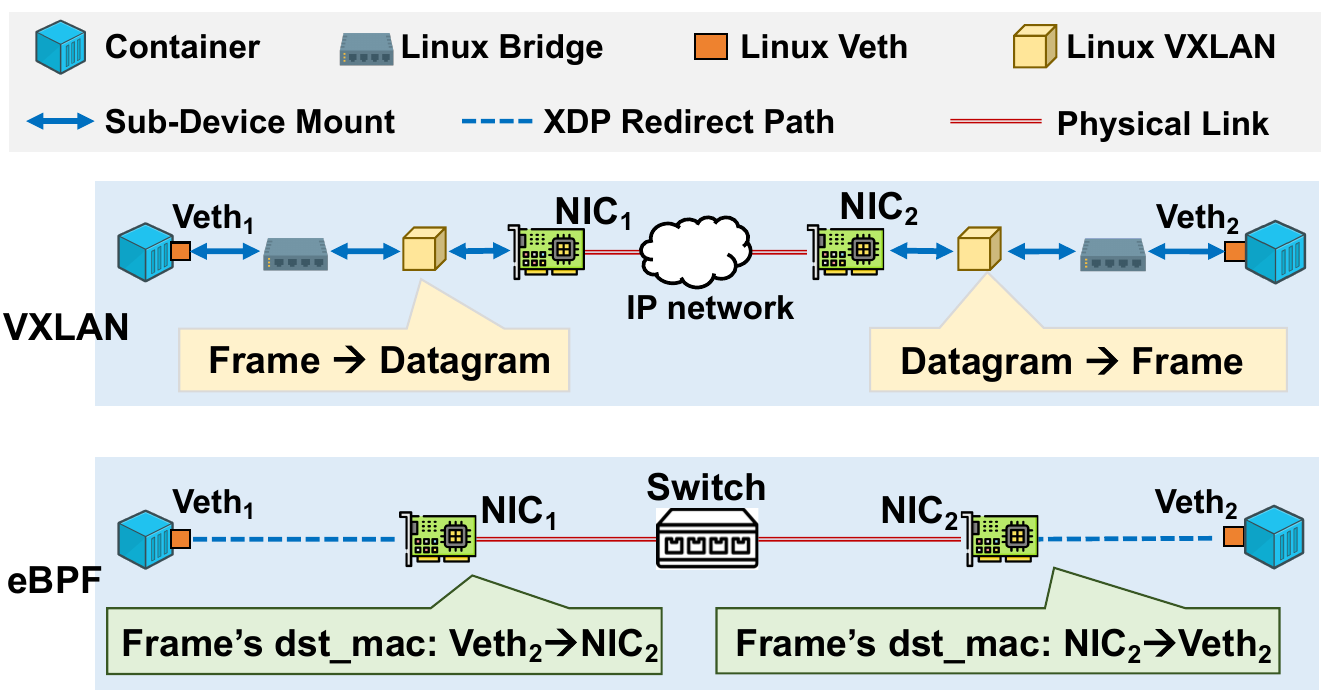}
\caption{Difference in data stream between inter-machine legacy links and inter-machine eBPF links}
\label{fig: eBPF_cross}
\end{figure}

Given the capability of multi-machine extension, the inter-machine links are used to transmit Ethernet frames to the correct container in the correct machine.
This type of link could be implemented via VXLAN or eBPF-based XDP.
OpenSN supports both methods.
We illustrate the difference based on Fig.~\ref{fig: eBPF_cross}.

\textbf{VXLAN:}
Ethernet frames will be transformed into UDP datagrams in the source container, and then sent to the destination machine.
After the UDP datagrams reach the destination machine, it will be transformed into Ethernet frames in VXLAN devices with specified VXLAN Network Identifier (VNI), which is then forwarded to the destination container.

\textbf{eBPF:}
For eBPF links, there is no need to conduct the transformation between Ethernet frames and datagrams.
Algorithm~\ref{alg: ebpf} shows procedure of the eBPF program attached to the virtual NIC.
Algorithm~\ref{alg: ebpf} consists of two phases.

\begin{itemize}
    \item Lines~\ref{Line: Update MAC Address}-\ref{Line: Update MAC Address End} update the destination MAC address of the packet. 
    If the field $\textit{Pkt.SrcMac}$ is in the key set of the $\textit{Map}_{\textit{dst}}$, then the forwarder will update the $\textit{Pkt.DstMac}$ to the corresponding value in the $\textit{Map}_{\textit{dst}}$.
    
    \item Lines~\ref{Line: eBPF Forward}-\ref{Line: eBPF Forward End} forward the packet based on three different scenarios. 
    If the field $\textit{Pkt.SrcMac}$ is in the key set of the $\textit{Map}_{\textit{fwd}}$, then this packet should be forwarded to a container inside the current machine. 
    Otherwise, the forwarder should check the direction of the packet. 
    If it is an $\textit{Egress}$ packet, then it should be sent out from the current machine.
    Otherwise, it is a packet for this machine, thus should be delivered to upper-layer protocols.    
\end{itemize}
When a packet is sent from a container instance, the eBPF program attached to the egress direction of a specified NIC in the source machine will modify the destination MAC address of the Ethernet frame.
Consequently, the frame will be forwarded to the NIC of the destination machine.
Accordingly, when a packet arrives at a machine's gateway NIC, the eBPF program attached to the ingress direction of the specified NIC in the destination machine will restore the received frame and redirect it to the destination container.

Note that the above procedure takes advantage of the point-to-point links in satellite networks, i.e., each source MAC address corresponds to a unique destination MAC address.

The eBPF-based implementation for inter-machine links provides the better function extensibility, since it does not rely on network-layer protocols in Linux kernel.
However, the VXLAN-based implementation relies on the UDP protocol and IP network.
Therefore, the eBPF-based implementation enables researchers to develop and evaluate novel network-layer architectures and protocols based on OpenSN.

\begin{algorithm}[t]
    \caption{Point-to-point Link eBPF Forwarder}
    \label{alg: ebpf}

    \KwIn{$\textit{Direction}$, $\textit{Pkt}$, $\textit{Map}_{\textit{fwd}}$, $\textit{Map}_{\textit{dst}}$, $\textit{Intf}_{\textit{egress}}$}
    
    \KwOut{$\textit{Target}$}

    \tcp{\hll{Update destination MAC address}}
    
    \If{$\textit{Pkt}.\textit{SrcMac} \in \textit{Map}_{\textit{dst}}$}{
    \label{Line: Update MAC Address}
    
        $\textit{Pkt}.{\textit{DstMac}} \leftarrow  \textit{Map}_{\textit{dst}}[\textit{Pkt}.\textit{SrcMac}]$
    }
    \label{Line: Update MAC Address End}

    \tcp{\hll{Forward Packet}}
    
    \If{$\textit{Pkt}.\textit{SrcMac} \in \textit{Map}_{\textit{fwd}}$}{
    \label{Line: eBPF Forward}
        $\textit{Target} \leftarrow \textit{Map}_{\textit{fwd}}[\textit{Pkt}.\textit{SrcMac}]$
    }
    \uElseIf{$\textit{Direction} = \textit{Egress}$}{
        $\textit{Target} \leftarrow \textit{Intf}_{\textit{egress}}$
    }
    \Else{
        $\textit{Target} \leftarrow  \textit{UP\_LAYER\_STACK} $
    }
    
    \textbf{return} $\textit{Target}$
    \label{Line: eBPF Forward End}
\end{algorithm}

\subsection{Enhanced Extensibility}
\label{Subsection: Extensibility}
Besides the above efficiency improvement, OpenSN also enhances the extensibility from the perspectives of architecture and container image.

\subsubsection{Architecture}
To facilitate the performance evaluation of new studies, SN emulators should exhibit good extensibility in terms of the emulation functionality.
The existing network simulation/emulation libraries usually enhance function extensibility by separating the business code (for simulation/emulation) from the operation code (of the simulation/emulation environment).
With the simulator/emulator SDKs, users can customize the simulation/emulation rules without understanding the underlying environment in detail.
OpenSN also follows such a design philosophy to achieve a better function extensibility.
As we mentioned in Fig.~\ref{fig: Architecture}, OpenSN users can flexibly specify the emulation parameters and emulation rules via the User-Defined Configurator.


\subsubsection{Container Images}
OpenSN also allows users to change the functionality of each node.
In general, SN emulation may involve many different types of nodes, e.g., satellites, GSs, terrestrial routers, switches, HTTP servers, and servers of different content providers.
Container technique provides the application-level isolation and the portability of virtual environment \cite{casalicchio2020state}.
With the help of Open Container Initiative (OCI)\cite{oci}, various applications can be deployed into the containers.
However, existing container-based emulators (e.g., StarryNet) usually support only a few images, and are hard to cover all possible configuration actions.
This is because they use direct function calls to configure the environment inside the container.
For example, the configurator of StarryNet generates the configuration file for BIRD routing software and launches the BIRD directly by itself.
Consequently, StarryNet can only use the image running BIRD routing software.
If one wants to use other images, then it is necessary to modify StarryNet and rewrite all that couples with this image.

To overcome the above drawback, OpenSN leaves the image configuration actions to the initialization program inside the container.
This way, OpenSN users can configure the environment inside the container by specifying necessary information in the initialization program. 
This approach provides users with better flexibility.
For example, if one would like to change the routing configuration, it only requires replacing the image and the initialization program inside the container, instead of modifying the entire emulator.
However, the aforementioned initialization program involves a technical problem, i.e., the 
initialization program cannot get information (e.g., addresses of other containers) outside the container. 
To address this issue, OpenSN implements the message forwarder to create an information path from User-Defined Configurator to the environment inside the container.
Based on User-Defined Configurator, OpenSN users can generate configuration data for each container and upload them to the Key-Value Database.
Upon detecting changes, the message forwarder will be noticed and forward them into the containers.
To sum up, OpenSN enables users to load any container image to the network node without changing the emulation library.

\subsubsection{Case Study}
We will demonstrate the function extensibility of OpenSN, and introduce how to build a video streaming scenario in satellite-terrestrial integrated network (STIN).
As shown in Fig.~\ref{fig: case_topology}, this scenario consists of three autonomous systems (ASes).
$\text{AS}_{1}$ corresponds to Starlink Shell-I constellation (i.e., 72 orbits and 22 satellites per orbit).
$\text{AS}_{2}$ consists of a terrestrial user (TU), a router, and a ground station (GS).
$\text{AS}_{3}$ consists of a content provider (CP), a router, and a GS.
The video stream is delivered from CP to TU via the satellite network.
OpenSN can build container images for different types of nodes in this scenario.
The detailed steps will be introduced in Section~\ref{Section: Usage}.

\section{OpenSN Usage}
\label{Section: Usage}
\begin{figure}
\centering
\includegraphics[width=0.95\linewidth]{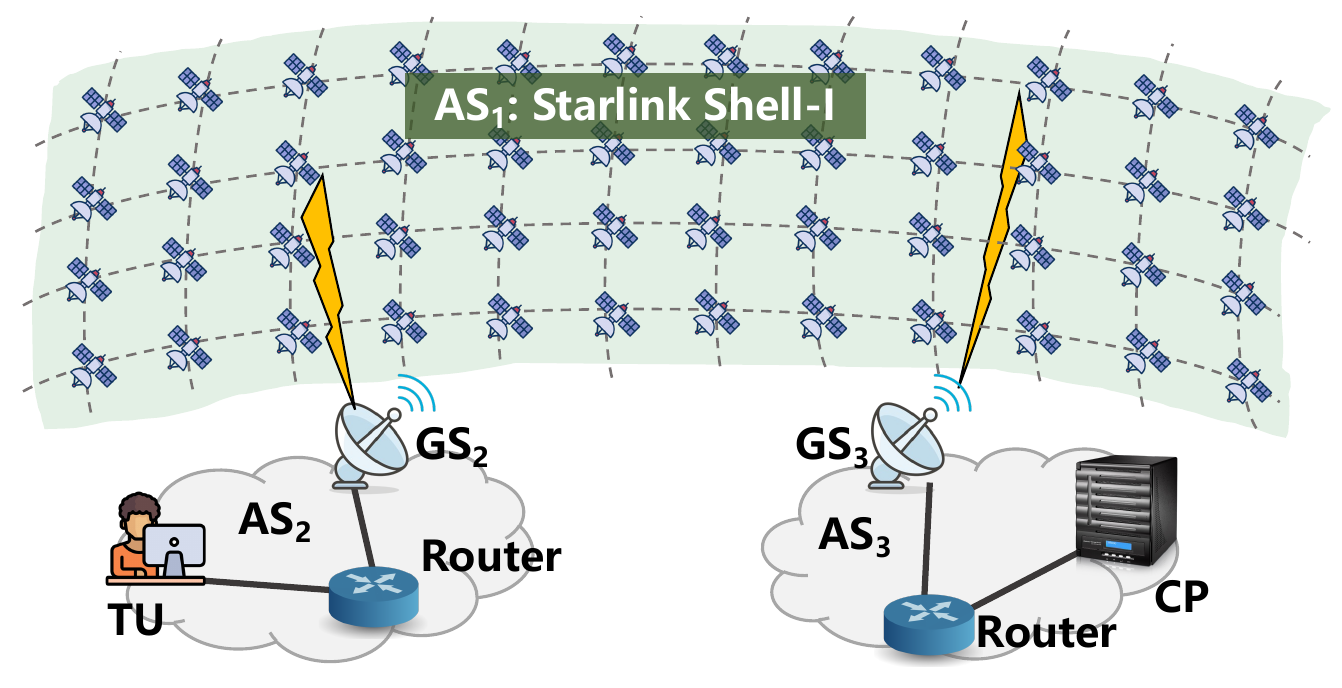}
\caption{Emulation scenario}
\label{fig: case_topology}
\end{figure}

This section introduces the usage of OpenSN.
Specifically, we will present the progress of establishing the emulation scenario shown in Fig.~\ref{fig: case_topology}.
The system requirement of deploying OpenSN is introduced in Section~\ref{Subsection: System Requirement}.
The illustration of OpenSN usage consists of container image configuration (in Section~\ref{Subsection: Container Image Configuration}), topology configuration (in Section~\ref{Subsection: topology configuation}), network emulation (in Section~\ref{Subsection: emulation and routing}). 
Finally, we introduce two representative scenarios of extending the emulation capability of OpenSN in Section~\ref{Subsection: Extend the emulation ability of OpenSN}.

\subsection{System Requirement}
\label{Subsection: System Requirement}
According to our experiments, the minimum requirement for a virtual machine is 1 vCPU and 2GB memory. 
Under this configuration, we successfully construct Iridium constellation (6 orbits with 11 satellites in each orbit) and deploy the OSPF routing protocol on each satellite.
This experiment demonstrates that OpenSN can function effectively under these modest hardware conditions. 
For larger constellations or more complex functionalities, additional resources will be necessary. 
Furthermore, we have tested OpenSN on Linux kernel versions ranging from 5.4 (Ubuntu 20.04) to 6.8 (Ubuntu 24.04). 
Given the emulator's design and the nature of standard Linux kernel updates, we expect OpenSN to be compatible with Linux kernel versions from 4.1 to the latest releases.

\subsection{Container Image Configuration}
\label{Subsection: Container Image Configuration}
Container image configuration of OpenSN consists of image creation and image assignment.

\subsubsection{Image Creation}
We first construct container images for the network nodes using Dockerfile \cite{dockerfile}.
Roughly speaking, there are two types of nodes in this emulation scenario.

\begin{figure}
\centering
\subfigure[Dockerfile for CP]{\label{fig: static-dockerfile-example}\includegraphics[width=0.48\linewidth]{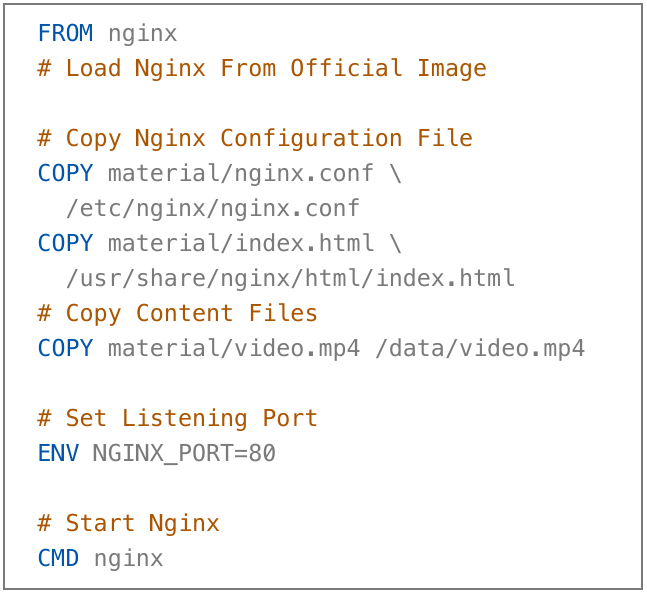}}
\subfigure[Dockerfile for satellite]{\label{fig: dockerfile-example}\includegraphics[width=0.48\linewidth]{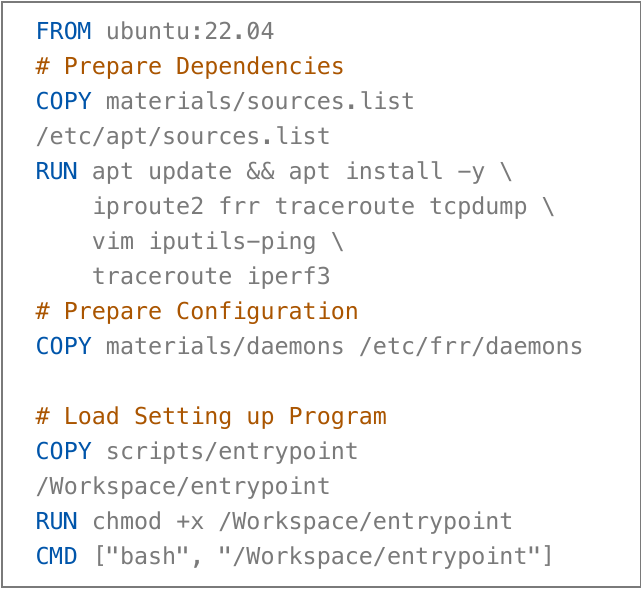}}
\subfigure[Setup program for satellite node]{\label{fig: dy-setting}\includegraphics[width=0.48\linewidth]{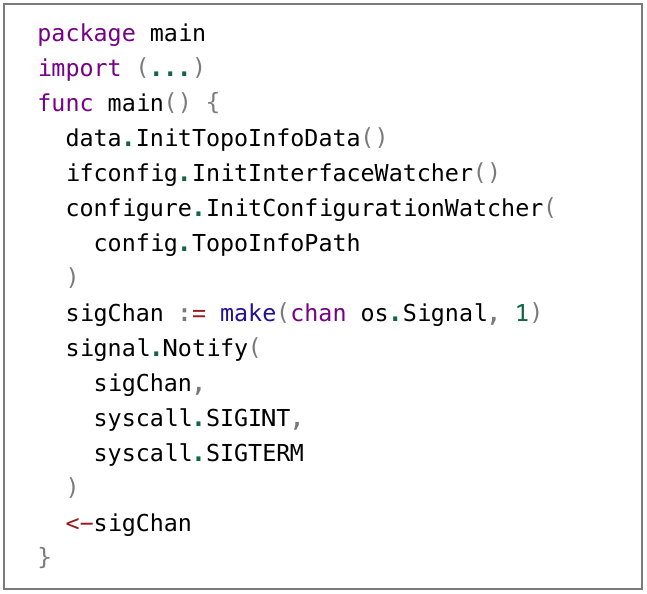}}
\subfigure[Container image assignment]{\label{fig: image-set}\includegraphics[width=0.48\linewidth]{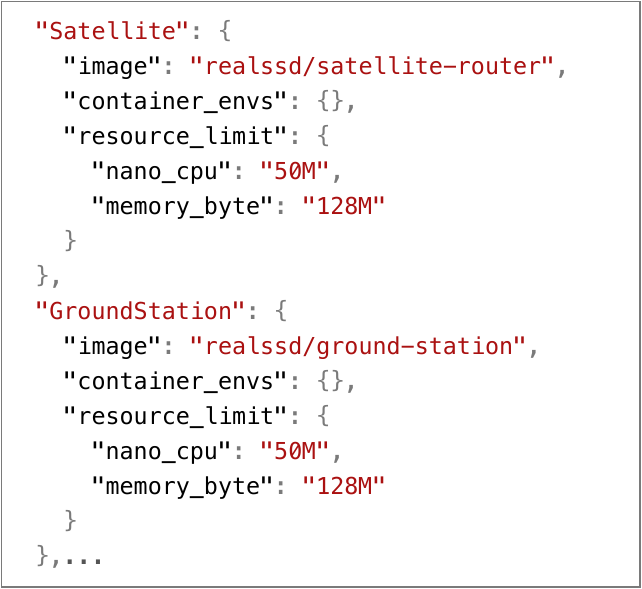}}
\caption{Container image configuration of OpenSN}
\label{fig: image-config}
\end{figure}

\textit{Static nodes could be pre-specified completely}. 
The routers and CP are static nodes in this scenario.
Fig.~\ref{fig: static-dockerfile-example} shows the Dockerfile of CP, which includes three parts: 1) Nginx application and libs from Nginx base image, 2) Nginx configuration files for this scenario, 3) resource contents (e.g., videos).

\textit{Dynamic nodes involve some time-varying status}, which cannot be specified before the emulation starts.
The satellites and TU are dynamic nodes in this scenario.
Fig.~\ref{fig: dockerfile-example} showcases the Dockerfile of a satellite node.
Besides regular static parameters, it also contains a user-defined setup program as the entry-point program.
Regarding the time-varying status (e.g., satellite positions and GSL endpoints), the dynamic node image requires an additional setup program in runtime dynamic configuration. 
Fig.~\ref{fig: dy-setting} shows the setup program of the satellite node.
At the beginning, the setup program waits for the topology data from User-Defined Configurator and initiates the route software with these data.
Upon receiving the topology data, the setup program will launch a network interface monitor and a topology configuration monitor to keep track of the topology changes.
Furthermore, the main thread of the setup program keeps sleeping until the container stops.

\subsubsection{Image Assignment}
\label{Subsection: apply-type-setting}
OpenSN could assign the created images to network nodes using a JSON file.
This file will be used in User-Defined Configurator.
Besides image assignment, one could also leverage the JSON file to configure the container environment (e.g., experiment timezone and default user directory), and specify the resource capacity for each type of node. 
Fig.~\ref{fig: image-set} showcases part of the JSON file regarding satellites and GSes. 
Specifically, the \textit{image} field specifies the container image.
The \textit{container\_envs} field is a map consisting of the environment variables that should be set into the containers.
The \textit{resource\_limit} field determines the resource limit of this type of container.

\subsection{Topology Configuration}
\label{Subsection: topology configuation}

\begin{figure}[t]
\centering
\includegraphics[width=0.9\linewidth]{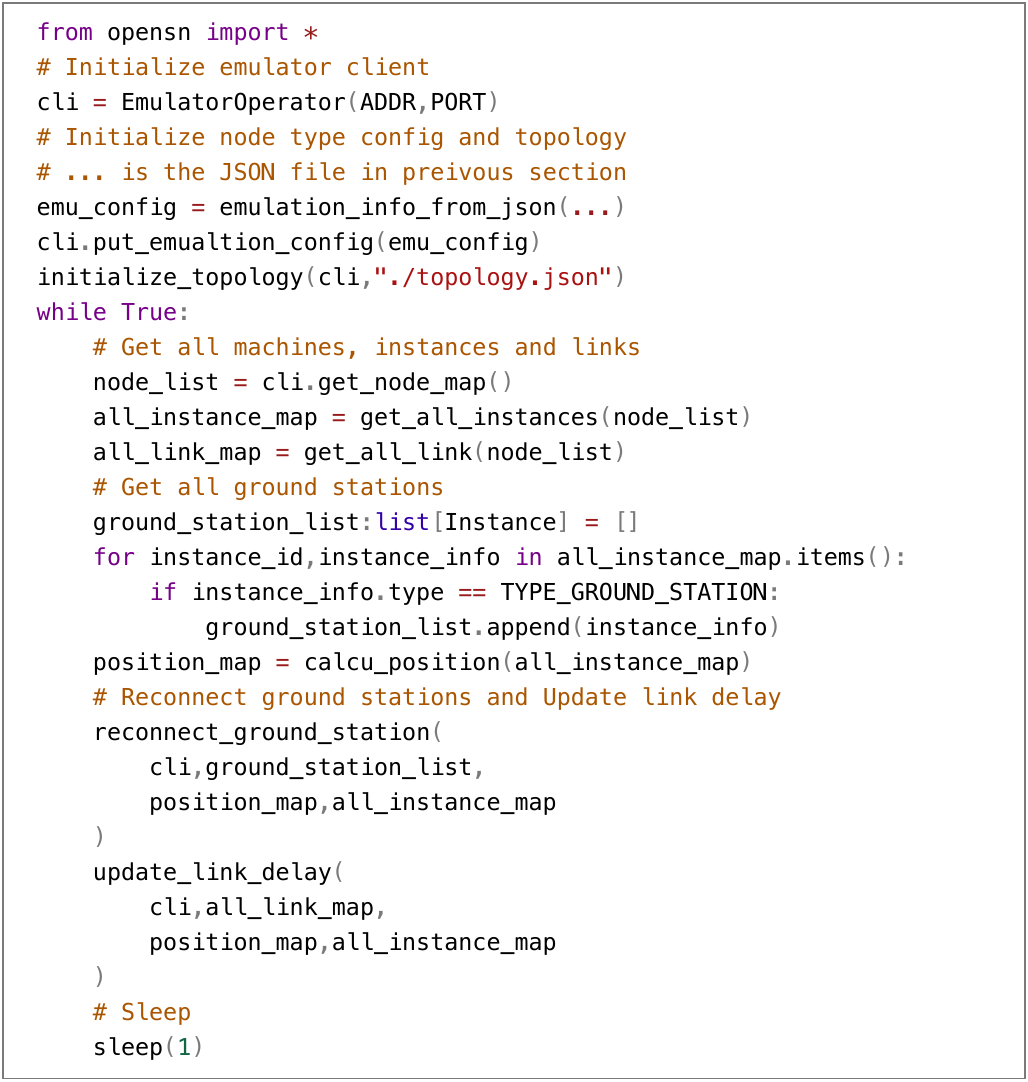}
\caption{User-Defined Configurator of the illustrative scenario}
\label{fig: topology-configurator}
\end{figure}

OpenSN controls the runtime topology via the User-Defined Configurator.
OpenSN provides SDK for Python to users to specify their topology control rules.
Specifically, OpenSN users first import this SDK, and then create an OpenSN operator to initialize the emulation environment. 
Fig.~\ref{fig: topology-configurator} shows the User-Defined Configurator for this emulation scenario.
In this scenario, we first create the initial topology based on the container image configuration introduced in Section~\ref{Subsection: apply-type-setting}.
Later on, User-Defined Configurator enters the regular emulation phase, during which the configurator synchronizes the topology state from Container Network Manager. 
Moreover, User-Defined Configurator calculates the real-time positions of satellites, and configures link handover and link delay update.

\subsection{Network Emulation}
\label{Subsection: emulation and routing}

With the above preparation, one could launch OpenSN's Container Network Managers via the following steps.

\subsubsection{Configure Container Network Manager}
\begin{figure}
\centering
\subfigure[Configure container network]{\label{fig: daemon-set}\includegraphics[width=0.48\linewidth]{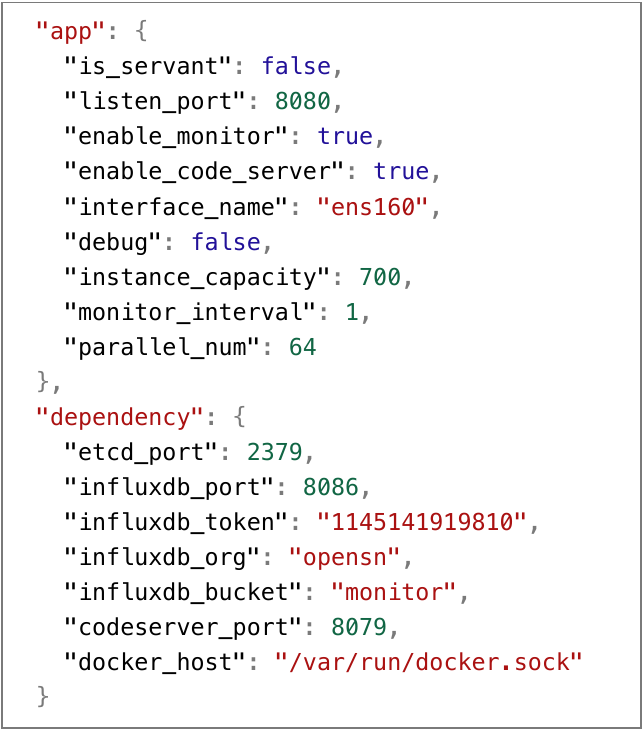}}
\subfigure[Launch container network]{\label{fig: daemon-start}\includegraphics[width=0.48\linewidth]{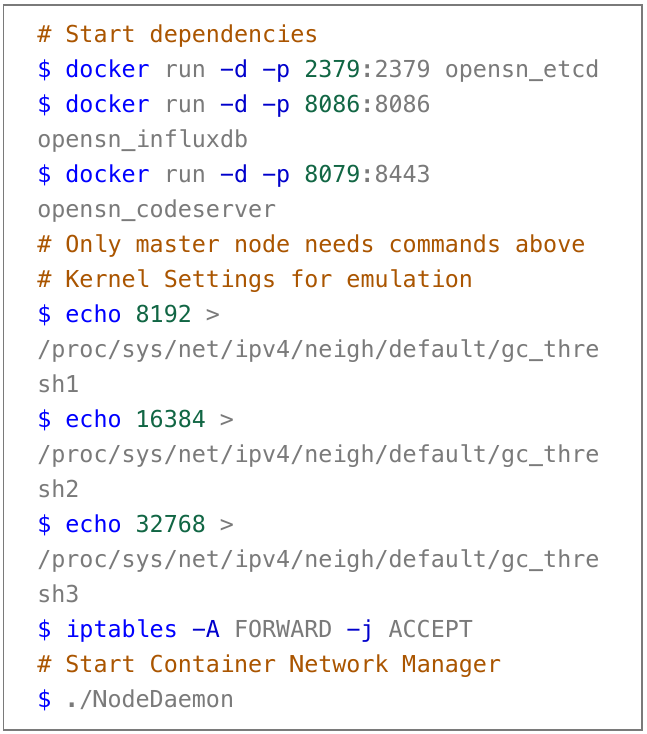}}
\caption{Launch OpenSN for SN emulation}
\label{fig: start opensn manager}
\end{figure}

OpenSN users need to edit the configuration file of Container Network Managers.
As shown in Fig.~\ref{fig: daemon-set}, the configuration file includes the following information. 
\begin{itemize}
\item The \textit{is\_servant} field indicates whether the Container Network Manager launches as a leader machine or a follower machine.
The cluster of the emulation environment needs only one leader machine.

\item The \textit{interface\_name} field indicates the gateway NIC for an emulation machine.
The address of the specified machine will be used in VXLAN device creation.
Moreover, the interface address will be used to exchange emulation data (e.g., emulation configuration and metrics data) with dependencies (e.g., Etcd and InfluxDB).

\item The \textit{instance\_capacity} field determines how many containers can be assigned to the machine. 
This field should be set according to the resource capacity of the deployed machine  (e.g., CPU and memory).

\item The \textit{parallel\_num} field determines the size of the coroutine pool.
The default value equals the number of processor cores.
\end{itemize}

\subsubsection{Launch Container Network Manager}
OpenSN users launch the Container Network Manager via the instructions shown in Fig.~\ref{fig: daemon-start}, which mainly consists of three parts:

\textbf{Step 1: Launch Dependency.}
Dependencies include all the assistive software (e.g., database) used in OpenSN Container Network Manager. 
Overall, OpenSN runs three dependencies: Etcd for data store, InfluxDB for metrics store, and CodeServer for conveniently editing User-Defined Configurator. 
Before launching OpenSN, one should launch the dependencies as Docker containers.

\textbf{Step 2: Update Kernel Setting.}
Note that launching OpenSN does not need to modify or recompile the kernel. 
Nevertheless, to emulate large-scale satellite constellations, two changes on the kernel settings are necessary.
First, one should remove the limitation on the garbage collection threshold of ARP cache (default is 128).
ARP table is used to map the IP address to the MAC address. 
The kernel stack uses this table to fill the MAC address field of the packet.
When the size of ARP cache reaches the threshold, the kernel will immediately start garbage collection and remove the old ARP items. 
However, large-scale constellation emulation generates a lot of ARP items, thus the limitation should be removed.
Second, one should enable the redirecting ability of bridge devices, which allows the frame to be transmitted via the virtual links.

\textbf{Step 3: Launch Container Network Manager.} 
Finally, one should launch OpenSN Container Network Manager via the main executable file named \textit{NodeDaemon}.

Note that the three steps above do not need to be conducted on each machine.
For the leader machine, the three steps are all necessary to launch Container Network Manager.
For follower machines, it is not necessary to launch dependencies, since the entire emulation cluster shares the dependency services deployed on the leader machine.

\subsubsection{Satellite Network Emulation}
After launching Container Network Manager, OpenSN users are ready for satellite network emulation.
The emulation environment involves the cooperation between User-Defined Configurator and the containers based on images.
Specifically, the emulation involves two phases.
In the first phase, the routers will exchange the routing information and generate the routing tables.
The emulation enters the second phase after routing convergence.
In this emulation scenario, TU sends requests to CP, which then delivers the video files to the TU.
During the emulation process, the real-time performance (e.g., throughput and delay) could be recorded and dumped into the file exchange directory (exposed to the host).
Accordingly, OpenSN users could obtain all the experimental data.

\subsection{Function Extensibility of OpenSN}
\label{Subsection: Extend the emulation ability of OpenSN}
OpenSN also provides users with the option to extend the routing protocols and network-layer protocol.

\subsubsection{Change Routing Protocols or Routing Software}
It is a common case to evaluate different routing protocols.
OpenSN allows users to use alternative routing protocols and develop new routing protocols based on the installed routing software.
First, OpenSN uses FRRouting~\cite{FRRouting} as the default routing software, which supports for many existing routing protocols. 
Users can change the routing protocols and modify the configuration file template in the container images.
Second, users can easily develop their new routing protocols based on FRRouting.
Third, OpenSN also allows users to use other routing software by building a new container image that loads the specific routing software.

\subsubsection{Switch from IPv4 to IPv6}
Suppose that an OpenSN user would like to use IPv6. 
The user only needs to make the following changes: a) change the address generation rule in the Topology Configurator; b) Change the routing protocol from OSPFv2 to OSPFv3 and modify the configuration template file in the container image.
The two steps above can be finished in the User-defined Topology Configurator of OpenSN.
They are separated from the emulation environment controller, thus it is convenient and efficient.

\begin{figure*}
	\centering
	\subfigure[SN construction]{\label{fig: container_construction}\includegraphics[height=0.19\linewidth]{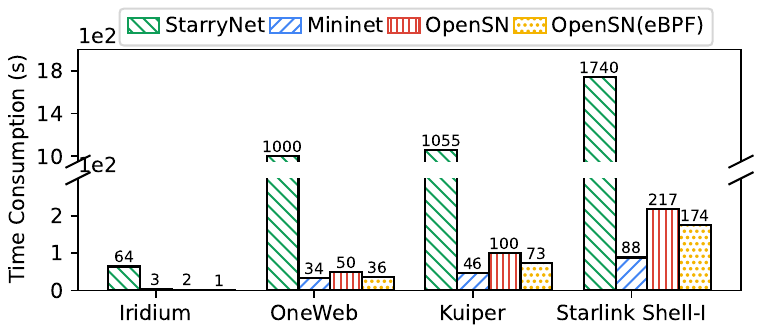}}\qquad
	\subfigure[SN deconstruction]{\label{fig: container_deconstruction}\includegraphics[height=0.19\linewidth]{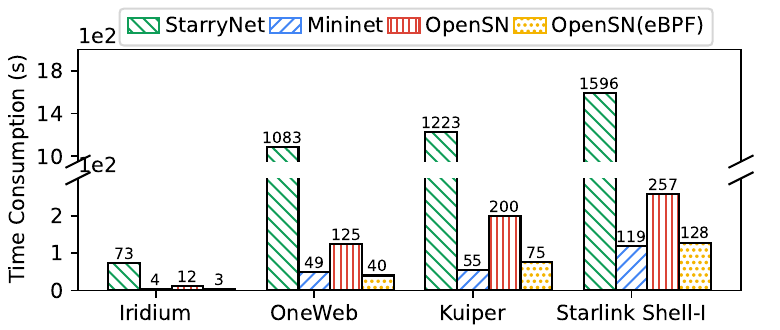}}
	\caption{Time consumption of constructing/deconstructing SN under OpenSN, StarryNet, and Mininet.}
	\label{fig: container_construction_deconstruction}
\end{figure*}

\section{Performance Evaluation}
\label{Section: Evaluation}

This section evaluates the performance of OpenSN in terms of emulation efficiency and system scalability.
Moreover, we will also compare it to existing open source SN emulators, i.e., StarryNet and LeoEM.
Recall that LeoEM is developed by adding trajectory and route calculation to Mininet, but does not allow for distributed routing software.
Hence the performance of LeoEM is similar to Mininet.
Moreover, the absence of distributed routing software makes LeoEM and Mininet more efficient than StarryNet and OpenSN (which run distributed routing software) in certain aspects.

\begin{table}[t]
\renewcommand{\arraystretch}{1.2}	
\centering	
\caption{LEO constellations used in our experiments} 
\label{Table: constellation}
\begin{tabular}{|c|c|c|c|c|c|c|c|}
\hline
\multirow{2}{*}{\makecell[c]{\textbf{Constellation}}} &
\multirow{2}{*}{\makecell[c]{\textbf{Altitude}}} & 
\multirow{2}{*}{\makecell[c]{\textbf{\# of} \\ \textbf{Orbits}}} & 
\multirow{2}{*}{\makecell[c]{\textbf{\# of} \\ \textbf{Satellites}}} & \multirow{2}{*}{\makecell[c]{\textbf{Inclination} \\ \textbf{Angle}}}\\ 
& & & & \\
\hline\hline
Iridium & 780 km   & 11 & 66 & 86.4$^\circ$ \\
\hline
OneWeb & 1200 km & 18 & 720 & 87.9 \\
\hline
Kuiper & 630 km & 34 & 1156 & 51.9$^\circ$\\
\hline
Starlink Shell-I & 550 km & 72 & 1584 & 53$^\circ$ \\
\hline
Starlink Shell-II & 540 km & 72 & 1584 & 53.2$^\circ$ \\
\hline
Starlink Shell-III & 570 km & 36 & 720 & 70$^\circ$ \\
\hline
Starlink Shell-IV & 560 km & 6 & 348 & 97.6$^\circ$ \\
\hline
Starlink Shell-V & 560 km & 4 & 172 & 97.6$^\circ$ \\
\hline
\end{tabular}
\end{table}

We build the experimental environment based on three DELL R7840 Servers with Intel(R) Xeon(R) Gold 5218 CPU and 191.5 GB memory in each.
Section~\ref{Subsection: SN Construction/deconstruction} introduces the performance of network construction/deconstruction.
Section~\ref{Subsection: Link State Update} introduces the network state update.
Section~\ref{Subsection: Runtime Resource Cost} introduces the runtime resource usage.
Section~\ref{Subsection: Impact of Resource Capacities on Emulation Performance} introduces the impact of resource capacity.
Section~\ref{Subsection: Impact of Machine Count on Emulation Performance} introduces the impact of the number of machines.
Section~\ref{Subsection: Five-shell Starlink Constellation Emulation} provides a case study on emulating the five-shells Stalink constellation.
Table~\ref{Table: constellation} shows the parameters of the LEO constellations used in our experiments.

\subsection{SN Construction/Deconstruction}
\label{Subsection: SN Construction/deconstruction}
We evaluate the efficiency of SN construction/deconstruction under OpenSN (without and with eBPF links), StarryNet, and LeoEM (or equivalently Mininet).
For OpenSN and StarryNet, we use a three-VM cluster with a total of 48 vCPU and 96GB memory.
For LeoEM (or equivalently Mininet), we use a single VM with a total of 48 vCPU and 96GB memory, since it does not support multi-machine extension.

\subsubsection{Network Construction/deconstruction}
For the large-scale LEO constellations, the time consumption of constructing and deconstructing the network is a nontrivial overhead, especially for container-based SN emulators running distributed routing software.
Therefore, we first evaluate the efficiency of network construction/deconstruction.

The two sub-figures in Fig.~\ref{fig: container_construction_deconstruction} show the time consumption of constructing and deconstructing SN, respectively.
Each sub-figure contains the results of four typical constellations, i.e., Iridium ($6\times11$), OneWeb ($18\times40$), Kuiper ($34\times34$), and Starlink Shell-I ($72\times22$).
Note that LeoEM (Mininet) achieves the best performance, since it does not run distributed routing software.
OpenSN slightly increases the construction/deconstruction time compared to LeoEM (Mininet).
Compared to the container-based StarryNet, OpenSN significantly reduces the network construction/deconstruction time (e.g., 6X-10X reduction) under typical LEO constellations. 
The construction time with eBPF is approximately 10\%-15\% faster than the classic virtual link implementation.
The destruction speed with eBPF links is up to 2x faster than with the classic virtual link.

\begin{figure*}
\centering
\subfigure[Construction progress under OneWeb]{\label{fig: construct_progress}\includegraphics[height=0.19\linewidth]{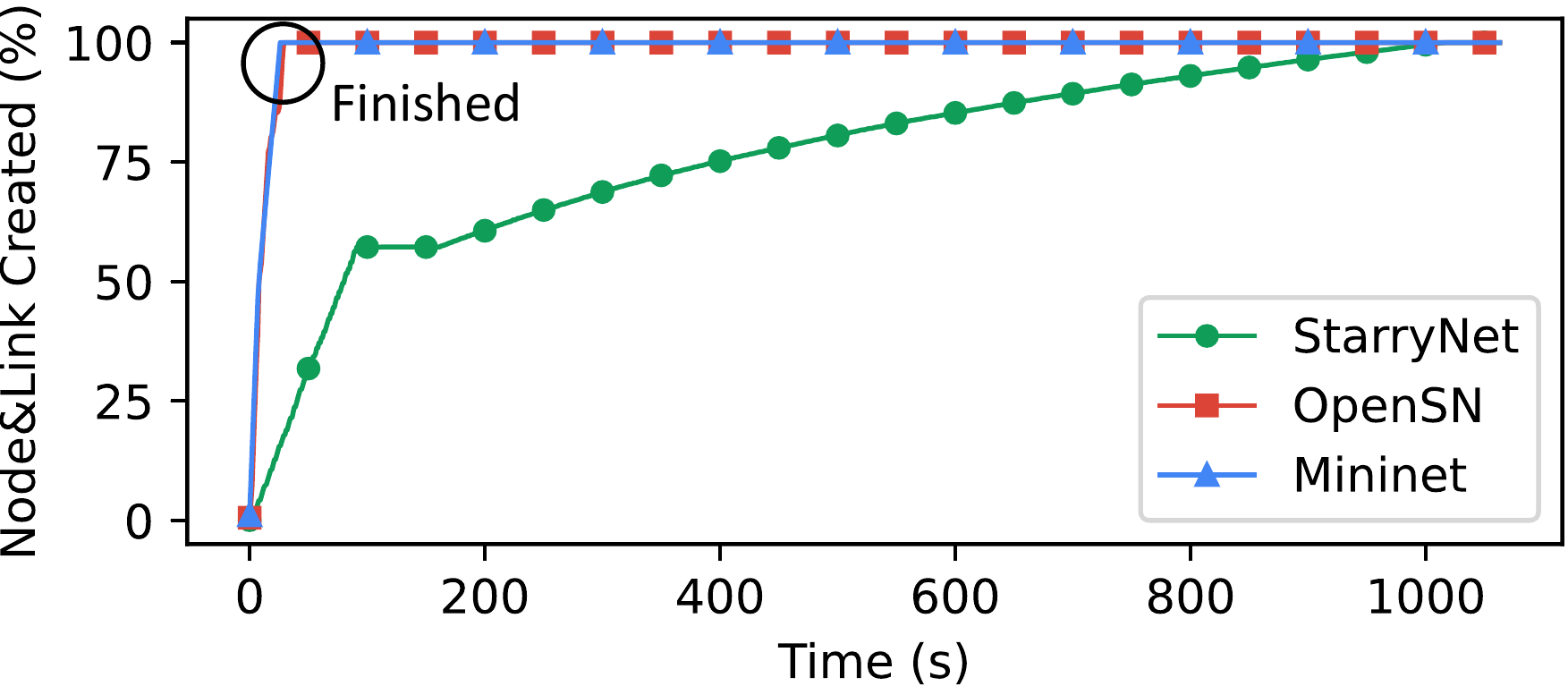}}\qquad
\subfigure[Deconstruction progress under OneWeb]{\label{fig: deconstruct_progress}\includegraphics[height=0.19\linewidth]{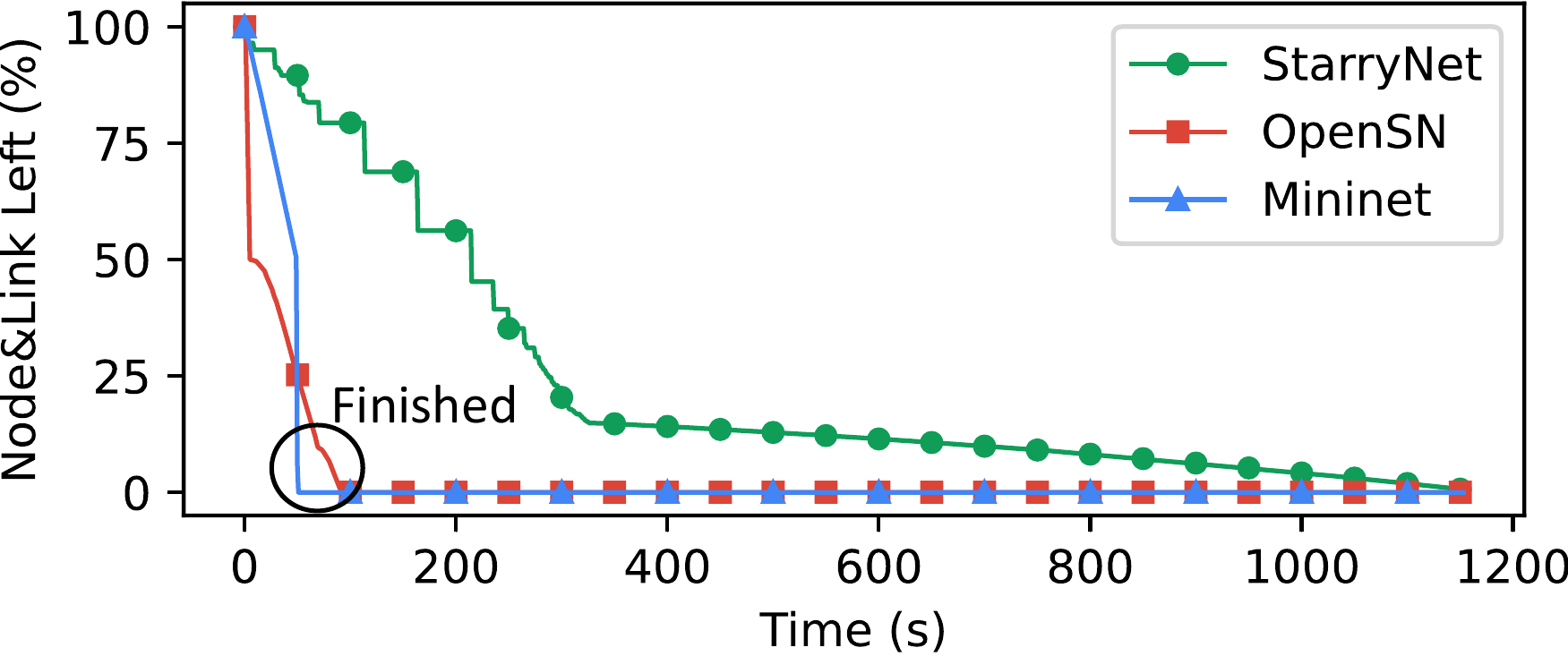}}
\caption{Progress of constructing/deconstructing SN under OpenSN, StarryNet, and Mininet.}
\label{fig: progress_construction_deconstruction}
\end{figure*}

\subsubsection{Progress of Network Construction/Deconstruction}
We take OneWeb as the example, and demonstrate the detailed progress of constructing/deconstructing SN under the three emulators.
Fig.~\ref{fig: progress_construction_deconstruction} shows the results.

Fig.~\ref{fig: construct_progress} plots the construction progress.
During 0-100s, StarryNet uses Docker Swarm to create containers in a parallel manner.
During 100-1000s, StarryNet is creating network links.
The construction speed during this period significantly decelerates due to two primary factors.
First, StarryNet uses Docker Network Controller to create links, which involves more actions.
Second, StarryNet adopts asynchronous interactions with Docker CLI, which generates many Docker client processes in operating system.
These processes consume significant computation resources and slow down the link creation speed.
The above issues have been addressed by the Virtual Link Manager of OpenSN.
Eventually, OpenSN achieves almost the same performance as Mininet in terms of the construction progress.

Fig.~\ref{fig: deconstruct_progress} plots the deconstruction progress.
Note that StarryNet is much slower than OpenSN and Mininet.
Specifically, during 0-300s, StarryNet uses threads and Docker CLI to delete containers in a parallel manner.
During 300-1150s, StarryNet is removing network links.
The deconstruction speed of this period significantly slows down for two reasons. 
First, StarryNet uses Docker CLI to remove containers and Docker Network Manager to remove links which involves more actions. 
Second, StarryNet schedules the container/link deletion tasks in a burst parallelism manner. 
The drawback of this scheduling policy is especially significant during the container deletion period (i.e., 0-300s). 
Note that the green curve is stair-wise during 0-300s.
That is, a couple of containers could be deleted together after a stasis (which lasts for about tens of seconds).
During each stasis, StarryNet executes many container deletion tasks instantaneously, which yields frequent process switches.
As mentioned in Section~\ref{Subsection: Network Emulation}, process switch is time-consuming, which reduces the efficiency of network deconstruction under StarryNet.
In contrast, OpenSN addresses this issue by limiting the number of tasks in execution state, which significantly speeds up network deconstruction.

\begin{figure*}
\centering
\subfigure[GSL connection update]{\label{fig: gsl-handover}\includegraphics[height=0.19\linewidth]{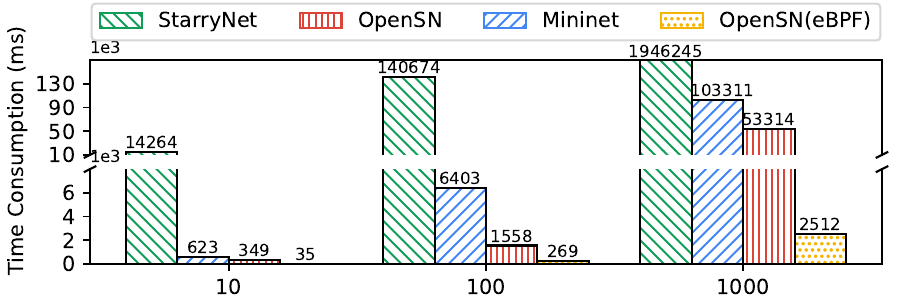}}\qquad
\subfigure[ISL latency update]{\label{fig:isl-update}\includegraphics[height=0.19\linewidth]{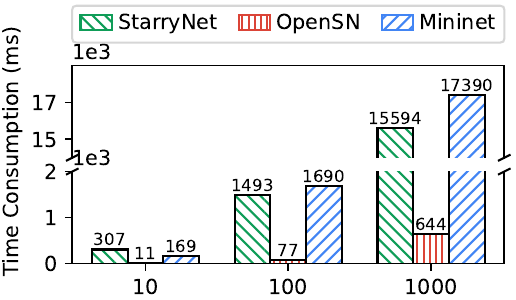}}
\caption{Efficiency of link state update under OpenSN, StarryNet and Mininet.}
\label{fig: change-response}
\end{figure*}

\subsection{Link State Update}
\label{Subsection: Link State Update}
The emulation of links is crucial for SN emulators.
Due to the mobile nature of LEO satellites, both GSLs and ISLs have time-varying states in terms of connectivity and propagation delay.
It will be a significant overhead to configure and update the link state for mega-constellations with thousands of nodes.
Next we evaluate the efficiency of executing operations of link state updates under OpenSN, StarryNet, and LeoEM (Mininet).
The configuration is the same as that in Section~\ref{Subsection: SN Construction/deconstruction}.
Specifically, we focus on GSL handover and ISL latency updates.
The two types of link-state changes are different in terms of virtual network emulation.
Specifically, GSL handover leads to topology changes. 
StarryNet and Mininet emulate topology changes by deleting and recreating the virtual network devices. 
Update of ISL latency only changes the parameters of links instead of deleting and recreating them.

Fig.~\ref{fig: gsl-handover} shows the time consumption of configuring GSL handovers. 
Overall, OpenSN significantly outperforms  StarryNet.
Moreover, OpenSN is $2\times$ faster than Mininet for configuring 10 GSL handovers and $4 \times$ faster for configuring 100 GSL handovers.
This is because OpenSN uses concurrency technology for virtual links.
Comparing the red and yellow bars shows that the eBPF link controller can further improve the efficiency of emulating GSL handover in OpenSN.
Specifically, OpenSN with eBPF links is $10\times$ faster than OpenSN with legacy links for configuring 10 GSL handovers, and $5 \times$ faster for configuring 100 GSL handovers. 
When the scale reaches 1000, the time consumption of OpenSN with legacy links increases greatly due to the concurrency bottleneck.
In contrast, OpenSN with eBPF links is capable of addressing this issue and keeps a low time consumption.

Fig.~\ref{fig:isl-update} shows the time consumption of configuring ISL latency updates.
Overall, OpenSN is more efficient than StarryNet and Mininet.
This is because OpenSN adopts concurrent execution and direct Netlink interaction. 
Both StarryNet and Mininet rely on command-line operations for updating link latency, which is less efficient than our adopted Netlink communication method in OpenSN. 

\begin{figure*}
\centering
\subfigure[OneWeb]{\label{fig: resource_oneweb}\includegraphics[height=0.205\linewidth]{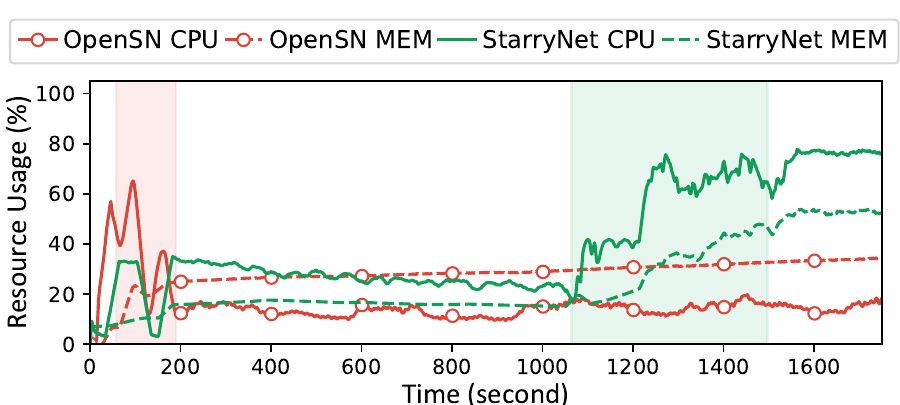}}\qquad
\subfigure[Starlink Shell-I]{\label{fig: resource_starlink}\includegraphics[height=0.205\linewidth]{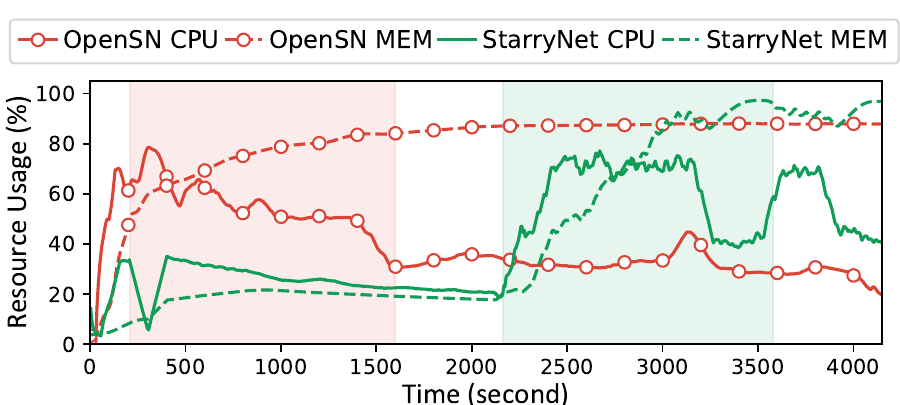}}
\caption{Resource usage across three periods: network construction, routing convergence (\textit{shaded}), and stable operation}
\label{fig: resource-cost}
\end{figure*}

\subsection{Runtime Resource Cost}
\label{Subsection: Runtime Resource Cost}
To obtain a better understanding of OpenSN, we investigate the runtime resource cost (i.e., CPU and memory).
In this experiment, we use a four-VM cluster with a total of 64 vCPU and 128GB memory for both OpenSN and StarryNet.
Fig.~\ref{fig: resource-cost} compares the real-time resource usage of OpenSN and StarryNet.
The two sub-figures correspond to emulating OneWeb constellation and Starlink Shell-I constellation.
Overall, there are three periods in each sub-figure, i.e., network construction, routing convergence, and stable operation.
The middle stage (i.e., routing convergence) is marked by red and green for OpenSN and StarryNet, respectively.
We have three remarks regarding the results in Fig.~\ref{fig: resource-cost}.

\begin{figure*}[t]
\centering
\subfigure[CPU/memory usage of Case A]{\label{fig: resource_oneweb_cmp_32c64g}\includegraphics[width=0.32\linewidth]{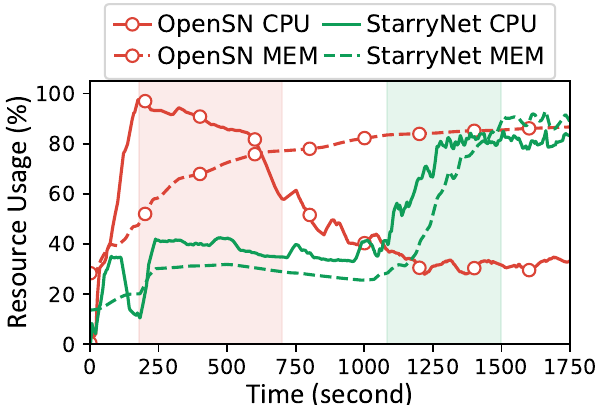}}
\subfigure[CPU/memory usage of Case B]{\label{fig: resource_oneweb_cmp_48c96g}\includegraphics[width=0.32\linewidth]{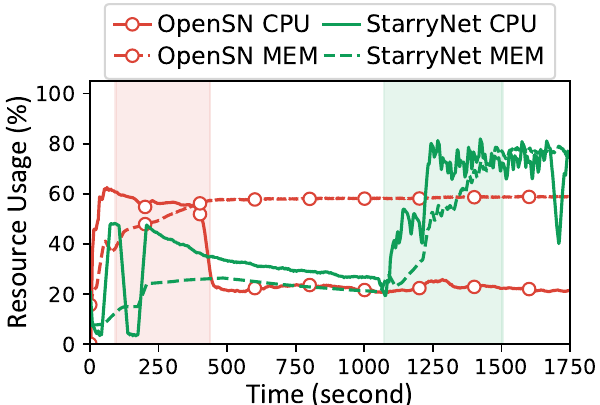}}
\subfigure[CPU/memory usage of Case C]{\label{fig: resource_oneweb_cmp_64c128g}\includegraphics[width=0.32\linewidth]{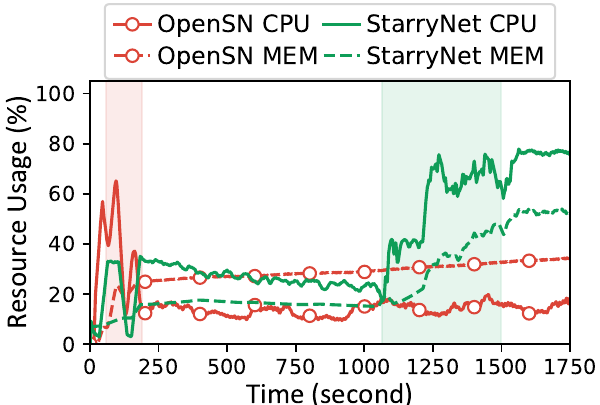}}
\caption{Resource usage under different configurations of resource}
\label{fig: different capacities}
\end{figure*}

\textbf{Remark 1:}
In each sub-figure, the red region is on the left-hand-side of the green region.
This means that OpenSN enters the stable operation period earlier than StarryNet.

\textbf{Remark 2:}
The experiments of emulating OneWeb and Starlink Shell-I are conducted under the same CPU and memory.
Hence it is a heavier workload to emulate Starlink Shell-I than OneWeb.
Comparing the two sub-figures also shows that the resource usage in Fig.~\ref{fig: resource_starlink} is higher than that in Fig.~\ref{fig: resource_oneweb}.

\textbf{Remark 3:}
Next we introduce the detailed resource usage of OpenSN and StarryNet during the three periods.
Specifically, we will focus on the heavy load case in Fig.~\ref{fig: resource_starlink}.
\begin{itemize}
\item \textbf{Network Construction} (\textit{before the shaded area}):
The CPU usage of OpenSN is rapidly increasing to 70\% during the network construction (i.e., 0-200s).
However, the CPU usage of StarryNet stays around 10\%-40\% during the network construction (i.e., 0-2100s).
That is, OpenSN makes full use of resources to speed up the construction progress of the container network.
The resource usage of StarryNet is low in this period, thus it takes a longer time to construct the same container network.

\item \textbf{Routing Convergence} (\textit{within the shaded area}):
The memory usage of both OpenSN and StarryNet is increasing during the routing convergence period.
This is due to the growth of routing table.
Furthermore, OpenSN's CPU usage is gradually decreasing (to 30\%) during the routing convergence period (i.e., red region).
In contrast, StarryNet's CPU usage is still at a high level (i.e., around 70\%) during the routing convergence period (i.e., green region).
This is because StarryNet takes the ping command as the daemon process of each container, which will frequently trigger soft interruptions.

\item \textbf{Stable Operation} (\textit{after the shaded area}):
During the stable operation period, the memory usage of OpenSN and StarryNet is similar.
Moreover, OpenSN incurs a lower CPU usage than StarryNet.
This is because OpenSN develops a lightweight daemon application for containers, which enters idle state after routing convergence.
This advantage allows OpenSN to maintain a more stable emulation environment for large-scale LEO constellations.
\end{itemize}

\subsection{Impact of Resource Capacity}
\label{Subsection: Impact of Resource Capacities on Emulation Performance}
We evaluate the runtime resource cost of emulating the same constellation under different resource configurations.
Specifically, we evaluate the performance of OpenSN under the following three configurations: 
Case A (32 vCPUs and 64 GB memory), Case B (48 vCPUs and 96 GB memory), and Case C (64 vCPUs and 128 GB memory).
Fig.~\ref{fig: different capacities} shows the results.
We have two observations.
First, comparing the width of red shaded regions in each sub-figure shows that the routing convergence time of OpenSN decreases as the resource capacity increases.
In contrast, comparing the width of green shaded regions in each sub-figure shows that the routing convergence time of StarryNet is almost the same when we increase the resource capacity.
Second, comparing the position of red shaded regions in each sub-figure shows that the time consumption of constructing OneWeb constellation on OpenSN decreases as the resource capacity increases.
In contrast, comparing the position of green shaded regions shows that the routing convergence time of StarryNet is almost the same when we increase the resource capacity.

\begin{figure*}[t]
\centering
\subfigure[One machine (32 vCPU \& 64G)]{\label{fig: one machine}\includegraphics[width=0.3\linewidth]{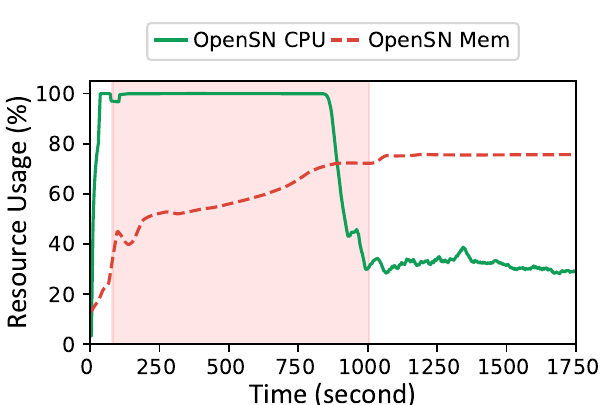}}\quad
\subfigure[Two machines (16 vGPU \& 32G for each)]{\label{fig: two machines}\includegraphics[width=0.3\linewidth]{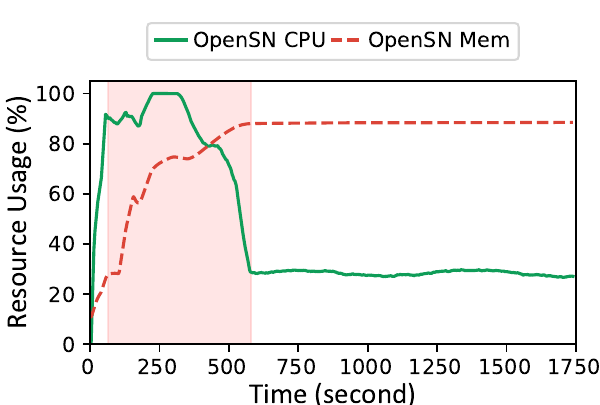}}\quad
\subfigure[Four machines (8 vCPU \& 16G for each)]{\label{fig: four machines}\includegraphics[width=0.3\linewidth]{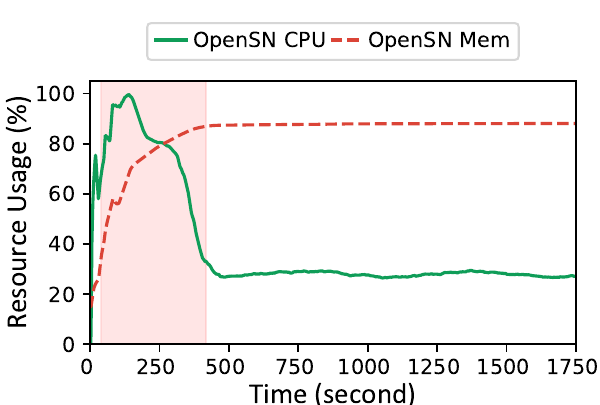}}
\caption{Resource usage of emulating OneWeb with OpenSN under different deployment specifications}
\label{fig: resource_oneweb_opensn}
\end{figure*}

\subsection{Impact of Number of Machines}
\label{Subsection: Impact of Machine Count on Emulation Performance}
\begin{figure}[t]
	\centering
	\includegraphics[height=0.45\linewidth]{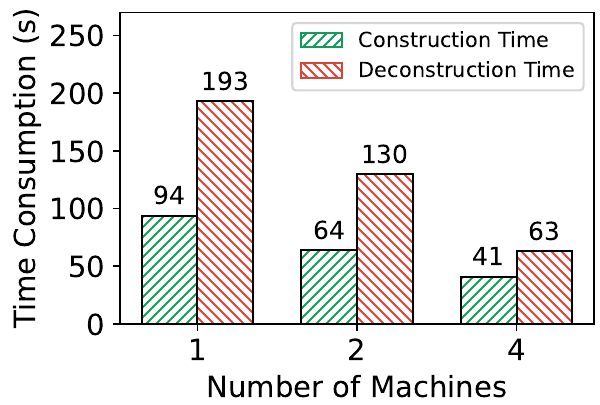}
	\caption{Impact of machine configurations}
	\label{fig: diff_node_time}
\end{figure}

To understand the importance of multi-machine emulation, we compare the performance of OpenSN under different number of machines.
We evaluate the efficiency of constructing/destructing OneWeb constellation under three configurations: 
Case I (one machine with 32 vCPUs and 64G memory), Case II (two machines with 16 vCPUs and 32G memory for each), and Case III (four machines with 8 vCPUs and 16G memory for each). 
Fig.~\ref{fig: resource_oneweb_opensn} shows the runtime resource consumption.
The three sub-figures correspond to the three cases.
In each sub-figure, there are three periods, i.e., network construction, routing convergence, and stable operation.
Comparing the shaded regions in three sub-figures shows that multi-machine deployment for OpenSN can speed up the construction of virtual network and the routing convergence.
Fig.~\ref{fig: diff_node_time} shows the time consumption of construction/destruction under different configurations.
The horizontal axis corresponds to three cases with different numbers of machines.
Note that as the number of machines increases, the time consumption of constructing/deconstructing OneWeb constellation decreases.

\begin{figure}[t]
    \centering
    \includegraphics[height=0.45\linewidth]{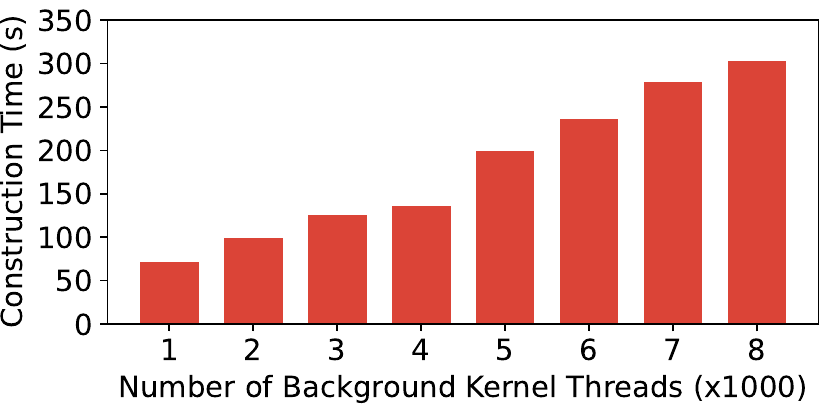}
    \caption{Impact of background kernel threads}
    \label{fig: cpu_many}
\end{figure}

To understand the above phenomenon, we further investigate the impact of existing kernel-threads on constellation construction efficiency.
Fig.~\ref{fig: cpu_many} shows the results.
The horizontal axis represents the number of existing kernel-threads before the construction of the OneWeb constellation.
The vertical axis represents the constellation construction time.
Note that the construction time increases as the number of kernel-threads increases. 
This observation indicates that the number of kernel threads existing in the system affects the speed of constellation construction, destruction, and related operations. 
If we deploy OpenSN on fewer machines, then there will be more kernel threads running on a single machine on average.
This will degrade the emulation efficiency.


\subsection{Case Study on Five-Shell Starlink Constellation}
\label{Subsection: Five-shell Starlink Constellation Emulation}
\begin{figure}
\centering
\includegraphics[height=0.45\linewidth]{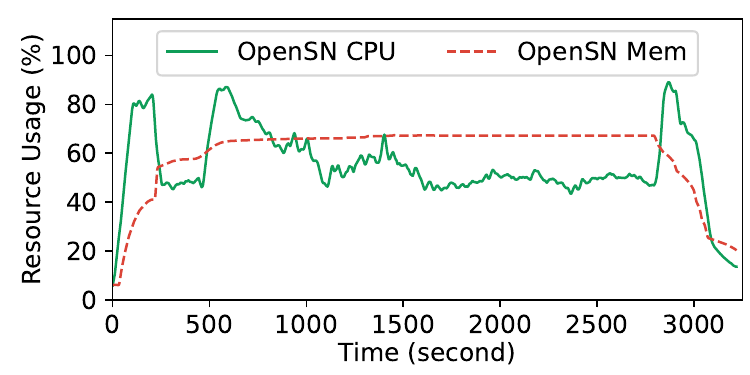}
\caption{Resource usage of emulating five-shell Starlink}
\label{fig: Starlink Shell 1-5 Process}
\end{figure}

To verify the scalability of OpenSN, we emulate the five-shell Starlink constellation shown in Table~\ref{Table: constellation}.
It poses significant challenges to emulate Starlink constellation with five shells due to our limited hardware resources (96 cores and 256 GB memory). 
To proceed, we divide each shell into small OSPF areas, and also separate the network construction process from the routing configuration process to avoid burst-like resource preemption.
Fig.~\ref{fig: Starlink Shell 1-5 Process} shows the results.

During the construction stage (i.e., 0-250s), OpenSN builds the topology of the container network.
During the configuration stage (i.e., 250-500s), OpenSN generates configuration files in each container, which primarily consumes IO bandwidth and results in moderate CPU usage.
Moreover, the routing convergence occurs independently in each shell.
The routing converges at the 600-th second for Shell V, at the 800-th second Shell IV, at the 900-th second for Shell III, at the 1500-th second for Shell II and Shell I.
After routing convergence, OpenSN enters the stable operation stage (1500-2800s). 
Shell IV and Shell V are the Walker-Star constellation with near-polar orbits, thus experience minor fluctuations in CPU usage due to ISL handover and routing re-convergence.
Finally, the emulation is terminated at the 2800-th second.

\section{Conclusion and Future Work}
\label{Section: Conclusion}
This paper presents OpenSN, i.e., an open source library for emulating large-scale LEO satellite networks.
OpenSN adopts container-based virtualization, thus allowing for running distributed software (e.g., routing) on each node, and achieves a better horizontal scalability than Mininet-based SN emulator LeoEM.
Compared to other container-based SN emulators (e.g., StarryNet), OpenSN streamlines the interaction with the Docker CLI and significantly reduces unnecessary operation during virtual link creation.
These modifications improve emulation efficiency and vertical scalability on a single machine.
Furthermore, OpenSN separates user-defined configuration from container network management via a Key-Value Database that records the necessary information for SN emulation.
Such a separation architecture enhances the function extensibility. 
Experimental results validate the performance of OpenSN.

The development of OpenSN is still in its early stage.
In the future, we would like to make the following improvements.
\begin{itemize}
\item \textit{Routing Protocols:}
The convergence time is a critical metric for routing protocols.
It is especially important for LEO satellite networks with topology dynamics.
In the future, we would like to develop more state-of-the-art routing protocols (e.g., OPSPF \cite{pan2019opspf} and LoFi \cite{shan2023routing}) on OpenSN, and then investigate the convergence time in LEO mega-constellations.

\item \textit{Networking Architectures:}
OpenSN is now built on host-centric IP networking architecture, which was initially designed for wired networks.
The topology characteristics of LEO constellations are quite different from those of wired networks.
To facilitate LEO satellite networking innovation, it is necessary to enrich the protocol stacks of OpenSN and make it support the emulation of alternative protocol stacks (e.g., content-centric NDN \cite{li2024istack} and link-centric LIPSIN \cite{jokela2009lipsin}) and even newly developed protocols.

\item \textit{Efficient Prototyping:}
The primary goal of network emulation is protocol prototypes.
Researchers and engineers face significant challenges on developing new network protocols, since they need to understand the kernel stack code and then develop the new protocols on top of it.
We will isolate the protocol logic from the kernel implementation for OpenSN, aiming to help expedite the development process and allow researchers to focus more on the protocol itself.
\end{itemize}

\bibliographystyle{IEEEtran}
\bibliography{ref}

\begin{thebibliography}{10}
\providecommand{\url}[1]{#1}
\csname url@samestyle\endcsname
\providecommand{\newblock}{\relax}
\providecommand{\bibinfo}[2]{#2}
\providecommand{\BIBentrySTDinterwordspacing}{\spaceskip=0pt\relax}
\providecommand{\BIBentryALTinterwordstretchfactor}{4}
\providecommand{\BIBentryALTinterwordspacing}{\spaceskip=\fontdimen2\font plus
\BIBentryALTinterwordstretchfactor\fontdimen3\font minus
  \fontdimen4\font\relax}
\providecommand{\BIBforeignlanguage}[2]{{%
\expandafter\ifx\csname l@#1\endcsname\relax
\typeout{** WARNING: IEEEtran.bst: No hyphenation pattern has been}%
\typeout{** loaded for the language `#1'. Using the pattern for}%
\typeout{** the default language instead.}%
\else
\language=\csname l@#1\endcsname
\fi
#2}}
\providecommand{\BIBdecl}{\relax}
\BIBdecl

\bibitem{apnet2024}
W.~Lu, Z.~Wang, Q.~Meng, and H.~Luo, ``Opensn: An open source library for
  emulating leo satellite networks,'' in \emph{Proceedings of Asia-Pacific
  Workshop on Networking (APNET)}, 2024.

\bibitem{IridiumISL}
{Iridium Next}, ``\url{https://www.iridium.com/network/}.''

\bibitem{StarlinkISL}
{Starlink Internet With Space Lasers},
  ``{https://cordcuttersnews.com/spacex-ramps-up-starlink-internet-speeds-with-thousands-of-space-lasers/}.''

\bibitem{handley2018delay}
M.~Handley, ``Delay is not an option: Low latency routing in space,'' in
  \emph{Proceedings of ACM Workshop on Hot Topics in Networks (HotNets)}, 2018,
  pp. 85--91.

\bibitem{bhattacherjee2018gearing}
D.~Bhattacherjee, W.~Aqeel, I.~N. Bozkurt, A.~Aguirre, B.~Chandrasekaran, P.~B.
  Godfrey, G.~Laughlin, B.~Maggs, and A.~Singla, ``Gearing up for the 21st
  century space race,'' in \emph{Proceedings of ACM Workshop on Hot Topics in
  Networks (HotNets)}, 2018, pp. 113--119.

\bibitem{handley2019using}
M.~Handley, ``Using ground relays for low-latency wide-area routing in
  megaconstellations,'' in \emph{Proceedings of ACM Workshop on Hot Topics in
  Networks (HotNets)}, 2019, pp. 125--132.

\bibitem{giuliari2020internet}
G.~Giuliari, T.~Klenze, M.~Legner, D.~Basin, A.~Perrig, and A.~Singla,
  ``{Internet} backbones in space,'' \emph{ACM SIGCOMM Computer Communication
  Review}, vol.~50, no.~1, pp. 25--37, 2020.

\bibitem{bhattacherjee2019network}
D.~Bhattacherjee and A.~Singla, ``Network topology design at 27,000 km/hour,''
  in \emph{Proceedings of International Conference on Emerging Networking
  Experiments And Technologies (CoNEXT)}, 2019, pp. 341--354.

\bibitem{zhang2022enabling}
Y.~Zhang, Q.~Wu, Z.~Lai, and H.~Li, ``Enabling low-latency-capable
  satellite-ground topology for emerging {LEO} satellite networks,'' in
  \emph{Proceedings of IEEE Conference on Computer Communications (INFOCOM)},
  2022, pp. 1329--1338.

\bibitem{li2021internet}
Y.~Li, H.~Li, L.~Liu, W.~Liu, J.~Liu, J.~Wu, Q.~Wu, J.~Liu, and Z.~Lai,
  ``{Internet} in space for terrestrial users via cyber-physical convergence,''
  in \emph{Proceedings of ACM Workshop on Hot Topics in Networks (HotNets)},
  2021, pp. 163--170.

\bibitem{chen2024shortest}
Q.~Chen, L.~Yang, Y.~Zhao, Y.~Wang, H.~Zhou, and X.~Chen, ``Shortest path in
  leo satellite constellation networks: An explicit analytic approach,''
  \emph{IEEE Journal on Selected Areas in Communications}, vol.~42, no.~5, pp.
  1175--1187, 2024.

\bibitem{pan2019opspf}
T.~Pan, T.~Huang, X.~Li, Y.~Chen, W.~Xue, and Y.~Liu, ``Opspf: Orbit prediction
  shortest path first routing for resilient leo satellite networks,'' in
  \emph{IEEE International Conference on Communications (ICC)}, 2019.

\bibitem{shan2023routing}
Q.~Shan, Z.~Wang, S.~Zhang, Q.~Meng, and H.~Luo, ``Routing in leo satellite
  networks: How many link-state updates do we need?'' in \emph{IEEE
  International Conference on Satellite Computing (Satellite)}, 2023.

\bibitem{taleb2008explicit}
T.~Taleb, D.~Mashimo, A.~Jamalipour, N.~Kato, and Y.~Nemoto, ``Explicit load
  balancing technique for ngeo satellite ip networks with on-board processing
  capabilities,'' \emph{IEEE/ACM transactions on Networking}, vol.~17, no.~1,
  pp. 281--293, 2008.

\bibitem{liang2021ndn}
T.~Liang, Z.~Xia, G.~Tang, Y.~Zhang, and B.~Zhang, ``Ndn in large leo satellite
  constellations: a case of consumer mobility support,'' in \emph{Proceedings
  of ACM Conference on Information-Centric Networking (ICN)}, 2021, pp. 1--12.

\bibitem{yan2024load}
F.~Yan, Z.~Wang, S.~Zhang, Q.~Meng, and H.~Luo, ``Load-aware hierarchical
  information-centric routing for large-scale leo satellite networks,'' in
  \emph{2024 IEEE Wireless Communications and Networking Conference
  (WCNC)}.\hskip 1em plus 0.5em minus 0.4em\relax IEEE, 2024, pp. 1--6.

\bibitem{xia2021adapting}
Z.~Xia, Y.~Zhang, T.~Liang, X.~Zhang, and B.~Fang, ``Adapting named data
  networking (ndn) for better consumer mobility support in leo satellite
  networks,'' in \emph{International ACM Conference on Modeling, Analysis and
  Simulation of Wireless and Mobile Systems}, 2021, pp. 207--216.

\bibitem{zhang2024link}
H.~Zhang, Z.~Wang, S.~Zhang, Q.~Meng, and H.~Luo, ``Link-identified routing
  architecture in space,'' \emph{IEEE Transactions on Network Science and
  Engineering}, 2024.

\bibitem{yan2024logic}
F.~Yan, Z.~Wang, S.~Zhang, Q.~Meng, and H.~Luo, ``Logic path identified
  hierarchical routing for large-scale leo satellite networks,'' \emph{IEEE
  Transactions on Network Science and Engineering}, 2024.

\bibitem{zhang2023link}
H.~Zhang, Z.~Wang, S.~Zhang, Q.~Meng, and H.~Luo, ``Optimizing link-identified
  forwarding framework in {LEO} satellite networks,'' in \emph{International
  Symposium on Modeling and Optimization in Mobile, Ad hoc, and Wireless
  Networks (WiOpt)}, 2023.

\bibitem{ekici2001network}
E.~Ekici, I.~F. Akyildiz, and M.~D. Bender, ``Network layer integration of
  terrestrial and satellite ip networks over bgp-s,'' in \emph{Proceedings of
  IEEE Global Telecommunications Conference (GLOBECOM)}, vol.~4.\hskip 1em plus
  0.5em minus 0.4em\relax IEEE, 2001, pp. 2698--2702.

\bibitem{zeng2024adaptive}
L.~Zeng, Z.~Wang, S.~Zhang, Q.~Meng, and H.~Luo, ``Adaptive inter-domain
  content retrieval in satellite-terrestrial integrated networks,'' in
  \emph{IEEE Global Communications Conference (GLOBECOM)}, 2024.

\bibitem{huang2024route}
S.~Huang, Z.~Wang, W.~Lu, K.~Shen, J.~Zhang, S.~Zhang, and H.~Luo, ``How to
  route cubic and bbr packets in space,'' in \emph{International Symposium on
  Modeling and Optimization in Mobile, Ad Hoc, and Wireless Networks
  (WiOpt)}.\hskip 1em plus 0.5em minus 0.4em\relax IEEE, 2024, pp. 265--272.

\bibitem{wang2024enabling}
Z.~Wang, X.~Lai, S.~Zhang, Q.~Meng, and H.~Luo, ``Enabling byzantine fault
  tolerance in access authentication for mega-constellations,'' \emph{IEEE/ACM
  Transactions on Networking}, 2024.

\bibitem{stk}
{Systems Tool Kit (STK)}, ``\url{https://www.agi.com/products/stk}.''

\bibitem{lai2020starperf}
Z.~Lai, H.~Li, and J.~Li, ``Starperf: Characterizing network performance for
  emerging mega-constellations,'' in \emph{Proceedings of IEEE International
  Conference on Network Protocols (ICNP)}, 2020.

\bibitem{kassing2020exploring}
S.~Kassing, D.~Bhattacherjee, A.~B. {\'A}guas, J.~E. Saethre, and A.~Singla,
  ``Exploring the internet from space with hypatia,'' in \emph{Proceedings of
  ACM Internet Measurement Conference (IMC)}, 2020.

\bibitem{yan2023comparative}
F.~Yan, H.~Luo, S.~Zhang, Z.~Wang, and P.~Lian, ``A comparative study of
  ip-based and icn-based link-state routing protocols in leo satellite
  networks,'' \emph{Peer-to-Peer Networking and Applications}, vol.~16, no.~6,
  pp. 3032--3046, 2023.

\bibitem{cheng2020comprehensive}
N.~Cheng, W.~Quan, W.~Shi, H.~Wu, Q.~Ye, H.~Zhou, W.~Zhuang, X.~Shen, and
  B.~Bai, ``A comprehensive simulation platform for space-air-ground integrated
  network,'' \emph{IEEE Wireless Communications}, vol.~27, no.~1, pp. 178--185,
  2020.

\bibitem{ns3}
{ns-3}, ``\url{https://www.nsnam.org}.''

\bibitem{omnet}
{OMNeT++}, ``\url{https://omnetpp.org}.''

\bibitem{mininet}
{Mininet}, ``\url{https://github.com/mininet/mininet}.''

\bibitem{docker}
{Docker}, ``\url{https://www.docker.com}.''

\bibitem{cao2023satcp}
X.~Cao and X.~Zhang, ``Satcp: Link-layer informed tcp adaptation for highly
  dynamic leo satellite networks,'' in \emph{Proceedings of IEEE Conference on
  Computer Communications (INFOCOM)}, 2023.

\bibitem{lai2023starrynet}
Z.~Lai, H.~Li, Y.~Deng, Q.~Wu, J.~Liu, Y.~Li, J.~Li, L.~Liu, W.~Liu, and J.~Wu,
  ``{StarryNet}: Empowering researchers to evaluate futuristic integrated space
  and terrestrial networks,'' in \emph{Proceedings of USENIX Symposium on
  Networked Systems Design and Implementation (NSDI)}, 2023.

\bibitem{pan2022docker}
T.~Pan, X.~Li, W.~Xue, Z.~Bian, T.~Huang, and Y.~Liu, ``{A Docker-based {LEO}
  Satellite Network Testbed},'' \emph{Chinese Journal of Computers}, vol.~45,
  no.~9, pp. 2029--2046, 2022.

\bibitem{mukerjee2020adapting}
M.~K. Mukerjee, C.~Canel, W.~Wang, D.~Kim, S.~Seshan, and A.~C. Snoeren,
  ``Adapting {TCP} for reconfigurable datacenter networks,'' in
  \emph{Proceedings of USENIX Symposium on Networked Systems Design and
  Implementation (NSDI)}, 2020.

\bibitem{pfandzelter2022celestial}
T.~Pfandzelter and D.~Bermbach, ``Celestial: Virtual software system testbeds
  for the leo edge,'' in \emph{Proceedings of the 23rd ACM/IFIP International
  Middleware Conference}, 2022, pp. 69--81.

\bibitem{agache2020firecracker}
A.~Agache, M.~Brooker, A.~Iordache, A.~Liguori, R.~Neugebauer, P.~Piwonka, and
  D.-M. Popa, ``Firecracker: Lightweight virtualization for serverless
  applications,'' in \emph{17th USENIX symposium on networked systems design
  and implementation (NSDI 20)}, 2020, pp. 419--434.

\bibitem{lai2021network}
J.~Lai, J.~Tian, K.~Zhang, Z.~Yang, and D.~Jiang, ``Network emulation as a
  service (neaas): Towards a cloud-based network emulation platform,''
  \emph{Mobile Networks and Applications}, vol.~26, pp. 766--780, 2021.

\bibitem{kivity2007kvm}
A.~Kivity, Y.~Kamay, D.~Laor, U.~Lublin, and A.~Liguori, ``kvm: the linux
  virtual machine monitor,'' in \emph{Proceedings of the Linux symposium},
  vol.~1, no.~8.\hskip 1em plus 0.5em minus 0.4em\relax Dttawa, Dntorio,
  Canada, 2007, pp. 225--230.

\bibitem{afhamisis2022testbed}
M.~Afhamisis, S.~Barillaro, and M.~R. Palattella, ``A testbed for lorawan
  satellite backhaul: Design principles and validation,'' in \emph{2022 IEEE
  International Conference on Communications Workshops (ICC Workshops)}.\hskip
  1em plus 0.5em minus 0.4em\relax IEEE, 2022, pp. 1171--1176.

\bibitem{LeoEMURL}
{LeoEM: a Real-Time LEO Satellite Network Emulator},
  ``\url{https://github.com/XuyangCaoUCSD/LeoEM}.''

\bibitem{Etalon}
{Etalon: A reconfigurable datacenter network emulator},
  ``\url{https://github.com/mukerjee/etalon}.''

\bibitem{bird}
{BIRD: Internet Routing Daemon}, ``\url{https://bird.network.cz}.''

\bibitem{StarryNetURL}
{StarryNet for the emulation of satellite Internet constellations},
  ``\url{https://github.com/SpaceNetLab/StarryNet}.''

\bibitem{quagga}
{Quagga}, ``\url{https://www.nongnu.org/quagga}.''

\bibitem{etcd}
{etcd}, ``\url{https://etcd.io}.''

\bibitem{mahalingam2014virtual}
M.~Mahalingam, D.~Dutt, K.~Duda, P.~Agarwal, L.~Kreeger, T.~Sridhar,
  M.~Bursell, and C.~Wright, ``Virtual extensible local area network (vxlan): A
  framework for overlaying virtualized layer 2 networks over layer 3
  networks,'' Tech. Rep., 2014.

\bibitem{vieira2020fast}
M.~A. Vieira, M.~S. Castanho, R.~D. Pac{\'\i}fico, E.~R. Santos, E.~P.~C.
  J{\'u}nior, and L.~F. Vieira, ``Fast packet processing with ebpf and xdp:
  Concepts, code, challenges, and applications,'' \emph{ACM Computing Surveys
  (CSUR)}, vol.~53, no.~1, pp. 1--36, 2020.

\bibitem{casalicchio2020state}
E.~Casalicchio and S.~Iannucci, ``The state-of-the-art in container
  technologies: Application, orchestration and security,'' \emph{Concurrency
  and Computation: Practice and Experience}, vol.~32, no.~17, p. e5668, 2020.

\bibitem{oci}
{Open Container Initiative}, ``\url{https://opencontainers.org}.''

\bibitem{dockerfile}
{Docker Docs}, ``\url{https://docs.docker.com/reference/dockerfile}.''

\bibitem{FRRouting}
{FRRouting}, ``\url{https://frrouting.org/}.''

\bibitem{li2024istack}
T.~Li, T.~Song, and Y.~Yang, ``{iStack}: A general and stateful name-based
  protocol stack for named data networking,'' in \emph{USENIX Symposium on
  Networked Systems Design and Implementation (NSDI)}, 2024, pp. 267--280.

\bibitem{jokela2009lipsin}
P.~Jokela, A.~Zahemszky, C.~Esteve~Rothenberg, S.~Arianfar, and P.~Nikander,
  ``Lipsin: Line speed publish/subscribe inter-networking,'' \emph{ACM SIGCOMM
  Computer Communication Review}, vol.~39, no.~4, pp. 195--206, 2009.

\end{thebibliography}

\begin{IEEEbiography}
[{\includegraphics[width=1in,height=1.25in,clip,keepaspectratio]{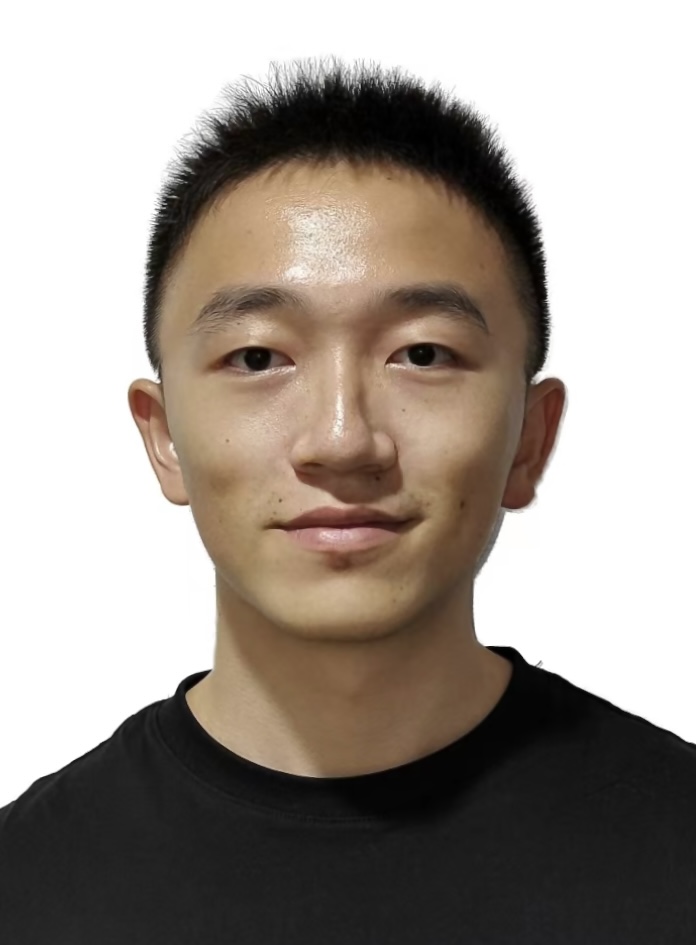}}]{Wenhao~Lu}
received the B.E. degree from the School of Computer Science and Engineering, Beihang University, Beijing, China, in 2024.
He is now working towards the Ph.D. degree in the School of Computer Science and Engineering, Beihang University, Beijing, China. 
His research interests include satellite networking and emulation platforms.
\end{IEEEbiography}

\begin{IEEEbiography}
[{\includegraphics[width=1in,height=1.25in,clip,keepaspectratio]{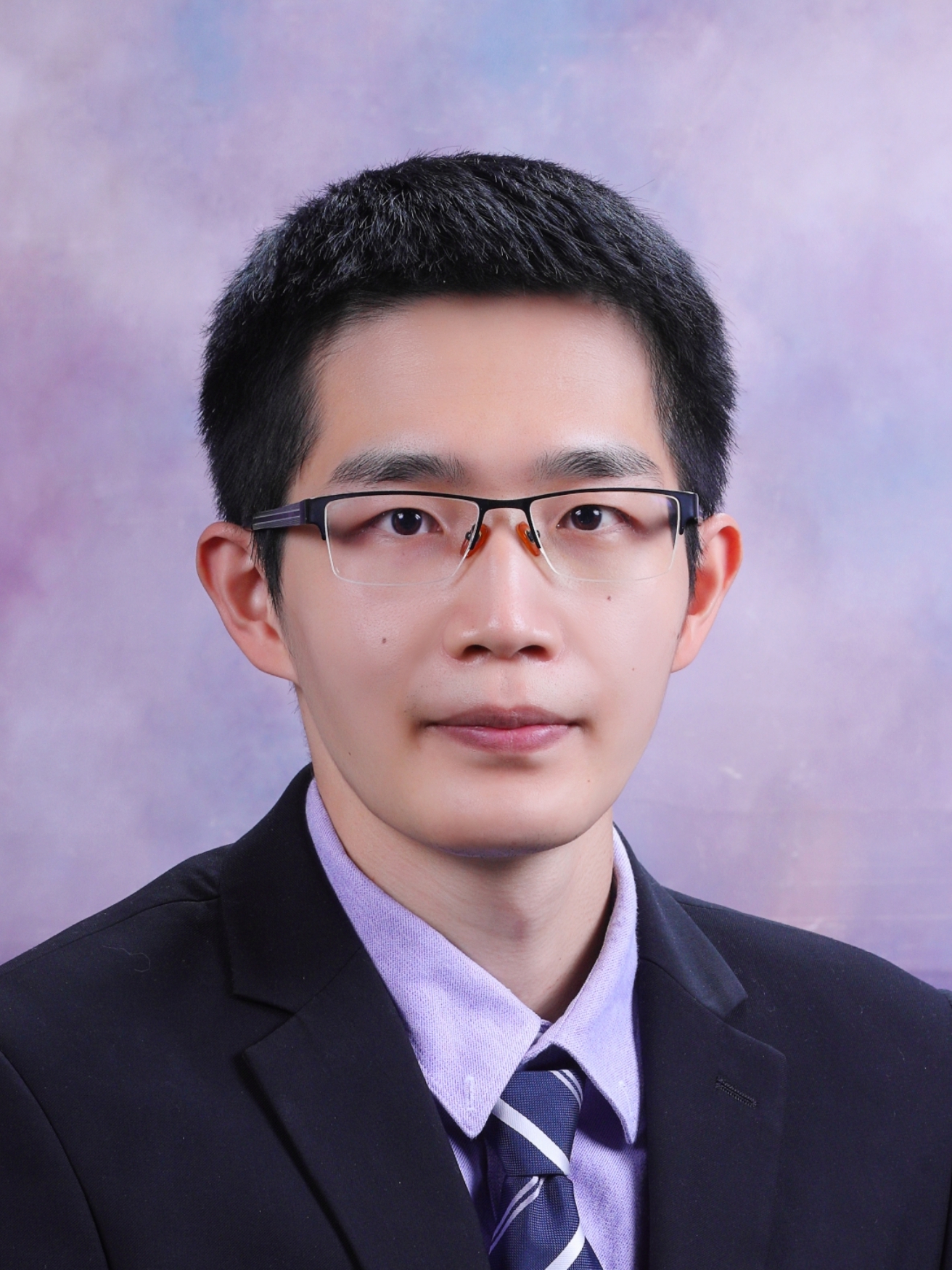}}]{Zhiyuan Wang} is an associate professor in School of Computer Science and Engineering, Beihang University.
He was a Post-Doctoral Fellow in Department of Computer Science and Engineering, The Chinese University of Hong Kong from 2019 to 2021.
He received his Ph.D. degree in Information Engineering, from The Chinese University of Hong Kong, in 2019. 
He received the B.Eng. degree in School of Information Science and Engineering, from Southeast University, Nanjing, in 2016. 
His research interest includes satellite-terrestrial integrated networks, edge/cloud computing, game theory, and online learning theory.
\end{IEEEbiography}

\begin{IEEEbiography}
[{\includegraphics[width=1in,height=1.25in,clip,keepaspectratio]{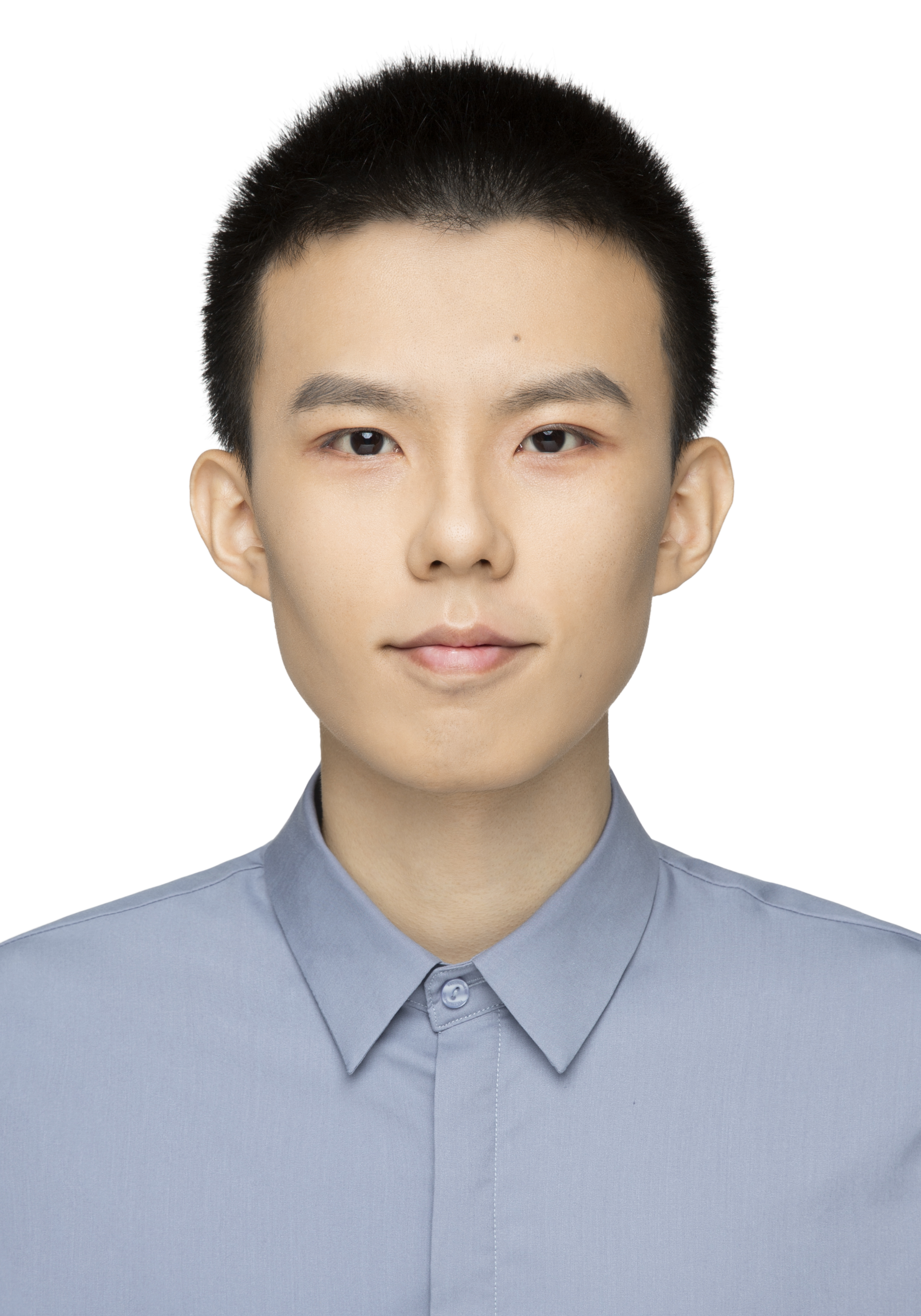}}] {Hefan Zhang} received the B.S. degree in School of Software Engineering from the Nanchang University, Nanchang, China, in 2022. 
He is currently pursuing the Ph.D. degree in the School of Computer Science and Engineering, Beihang University. His research interests include Internet architecture, Satellite networks routing, and Integration of satellite-terrestrial networks.
\end{IEEEbiography}

\begin{IEEEbiography}
[{\includegraphics[width=1in,height=1.25in,clip,keepaspectratio]{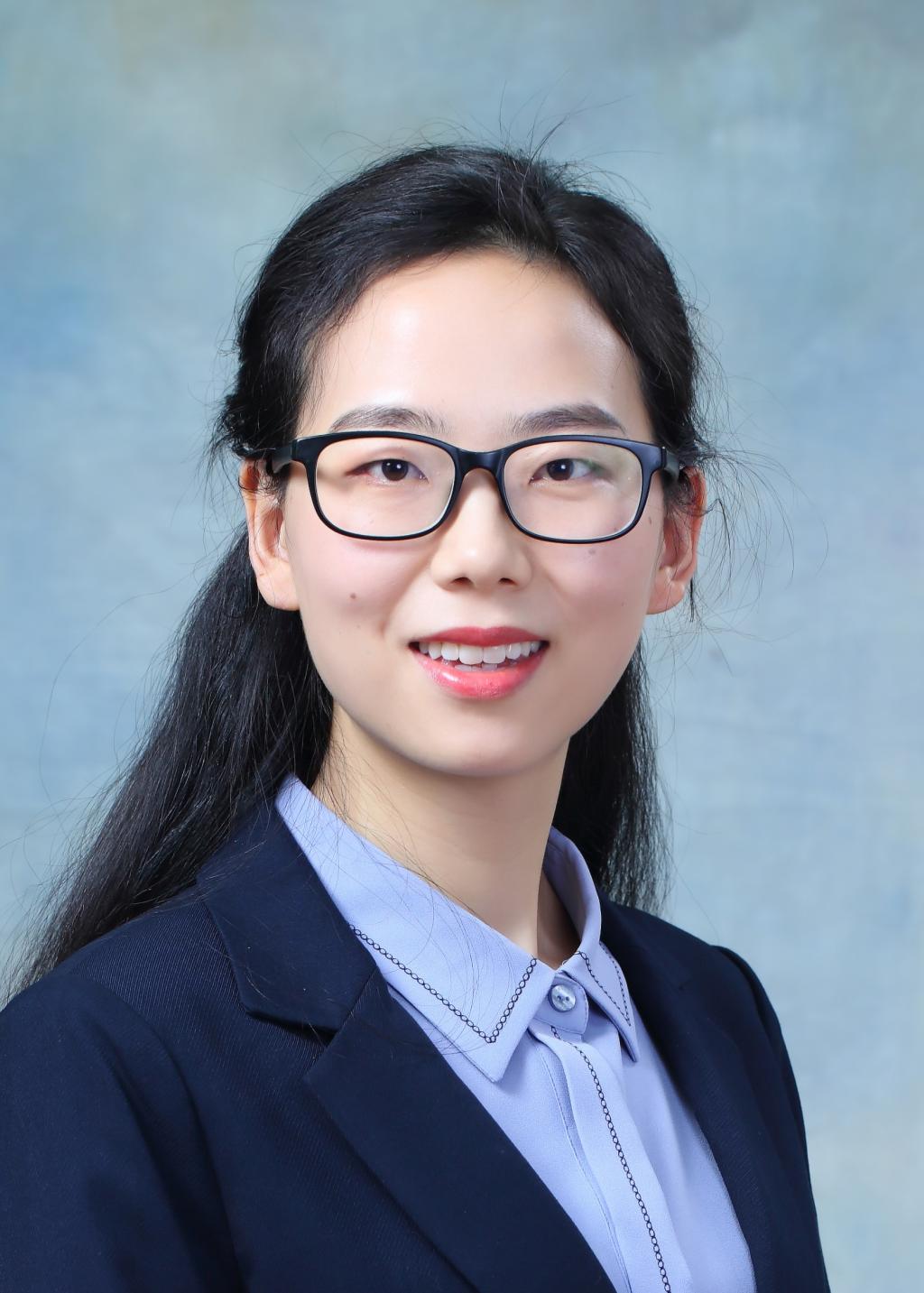}}]
{Shan Zhang} (Member, IEEE) received the Ph.D. degree in electronic engineering from Tsinghua University, Beijing, China, in 2016. She is currently an associate professor at the School of Computer Science and Engineering, Beihang University, Beijing, China. 
She was a postdoctoral fellow in the Department of Electronical and Computer Engineering, University of Waterloo, Ontario, Canada, from 2016 to 2017. 
Her research interests include mobile edge computing, wireless network virtualization, and intelligent management. 
She received the Best Paper Award at the Asia-Pacific Conference on Communication, in 2013. 
She has been serving as an associate editor for Peer-to-Peer Networking and Applications, and a guest editor for China Communications. 
\end{IEEEbiography}

\begin{IEEEbiography}
[{\includegraphics[width=1in,height=1.25in,clip,keepaspectratio]{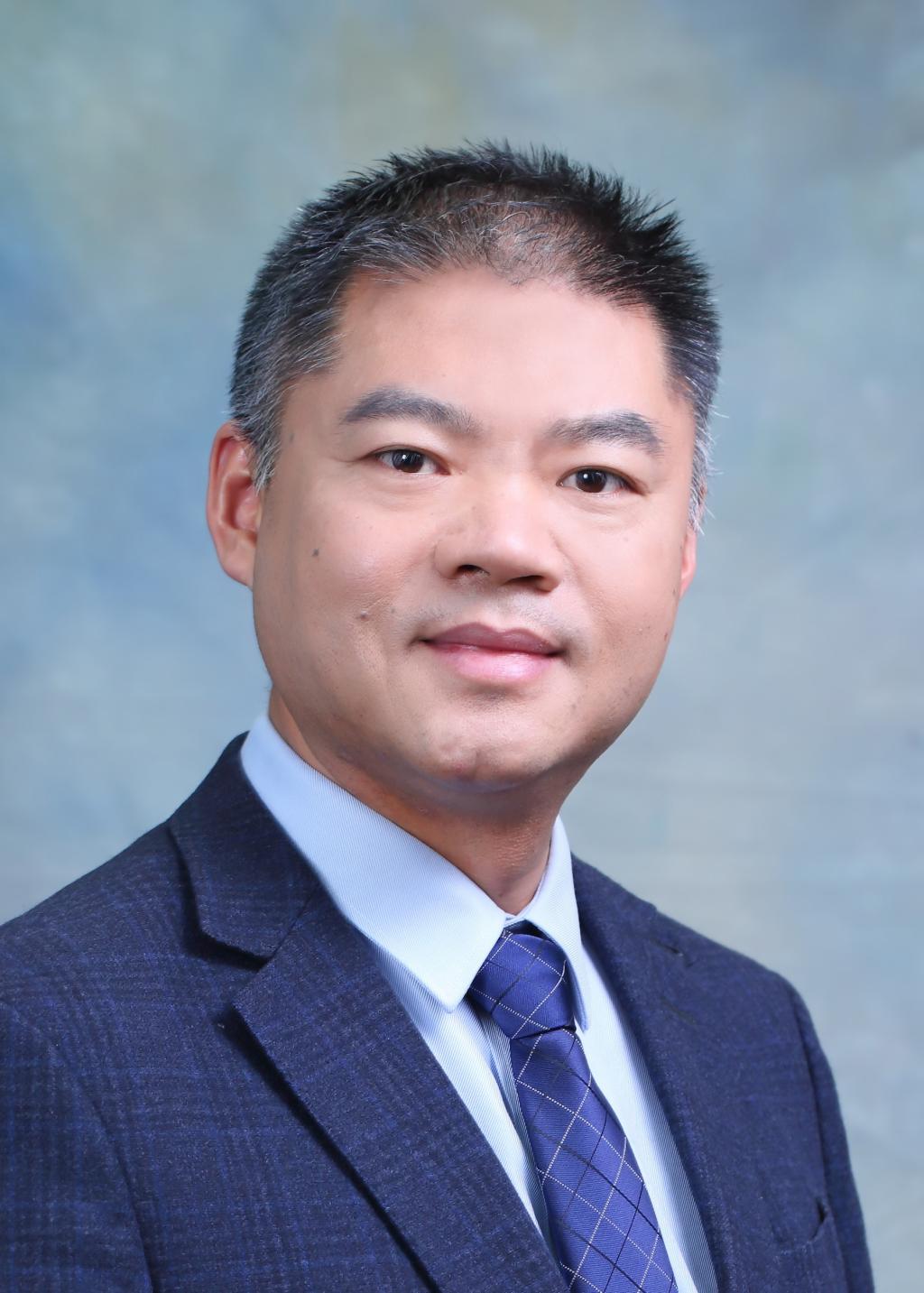}}]
{Hongbin Luo} (Member, IEEE)
received the B.S. degree from Beihang University, in 1999, and the M.S. (with honors) and Ph.D. degrees in communications and information science from the University of Electronic Science and Technology of China (UESTC), in June 2004 and March 2007, respectively. He is currently a professor at the School of Computer Science and Engineering, Beihang University. From June 2007 to March
2017, he worked at School of Electronic and Information Engineering, Beijing Jiaotong University.
From September 2009 to September 2010, he was a visiting scholar at Department of Computer Science, Purdue University. 
He has authored more than 50 peer-reviewed papers in international journals and conference proceedings. 
His research interests include network architecture, routing, and traffic engineering.
\end{IEEEbiography}

\end{document}